\documentclass[pre,tighten,eqsecnum]{revtex4}

\usepackage{amssymb,amsmath}

\usepackage{graphicx}

\usepackage{color}

\newcommand\beq{\begin{equation}}
\newcommand\eeq{\end{equation}}
\newcommand\beqa{\begin{eqnarray}}
\newcommand\eeqa{\end{eqnarray}}
\newcommand{\dd}{\text{d}}

\newcommand{\al}{\alpha}

\begin{document}

\title{Instabilities in granular gas-solid flows}
\author{Rub\'en G\'omez Gonz\'alez}
\affiliation{Departamento de F\'{\i}sica, Universidad de Extremadura, Universidad de Extremadura, E-06071 Badajoz, Spain}
\author{Vicente Garz\'o}
\email{vicenteg@unex.es} \homepage{http://www.eweb.unex.es/eweb/fisteor/vicente/}
\affiliation{Departamento de F\'{\i}sica and Instituto de Computaci\'on Cient\'{\i}fica Avanzada (ICCAEx), Universidad de Extremadura, E-06071 Badajoz, Spain}

\begin{abstract}

A linear stability analysis of the hydrodynamic equations with respect to the homogeneous cooling state is performed to study the conditions for stability of a suspension of solid particles immersed in a viscous gas. The dissipation in such systems arises from two different sources: inelasticity in particle collisions and viscous friction dissipation due to the influence of the gas phase on the solid particles. The starting point is a suspension model based on the (inelastic) Enskog kinetic equation. The effect of the interstitial gas phase on the dynamics of grains is modeled though a viscous drag force. The study is carried out in two different steps. First, the transport coefficients of the system are obtained by solving the Enskog equation by means of the Chapman-Enskog method up to first order in spatial gradients. Explicit expressions for the Navier-Stokes transport coefficients are obtained in terms of the volume fraction, the coefficient of restitution and the friction coefficient characterizing the amplitude of the external force. Once the transport properties are known, then the corresponding linearized hydrodynamic equations are solved to get the dispersion relations. In contrast to previous studies [V. Garz\'o \emph{et al.}, Phys. Rev. E \textbf{93}, 012905 (2016)], the hydrodynamic modes are \emph{analytically} obtained as functions of the parameter space of the system. For a $d$-dimensional system, as expected linear stability shows $d-1$ transversal (shear) modes and a longitudinal ``heat'' mode to be unstable with respect to long enough wavelength excitations. The results also show that the main effect of the gas phase is to decrease the value of the critical length $L_c$ (beyond which the system becomes unstable) with respect to its value for a dry granular fluid. Comparison with direct numerical simulations for $L_c$ shows a qualitative good agreement for conditions of practical interest.\\


\end{abstract}

\draft
\date{\today}
\maketitle

\section{Introduction}
\label{sec1}

The simplest way of capturing the dynamics of granular media under rapid flow conditions is through a fluid of smooth hard spheres with inelastic collisions \cite{G03}. The inelasticity of collisions is accounted for by a (positive) constant coefficient of normal restitution $\al \leqslant 1$ only affecting the translational degrees of freedom of grains. In spite of its simplicity, the model has been shown to be reliable for describing properties of granular flows especially those related with collisional dissipation. Unlike molecular fluids, granular fluids show spontaneous formation of velocity vortices and density clusters (microstructures characterized by regions of highly concentrated particles) in freely cooling flows. The origin of this type of instability is associated with the dissipative nature of collisions and is likely the
most characteristic property that makes granular flows so distinct from ordinary fluids. Detected first by Goldhirsch and Zanetti \cite{GZ93} and McNamara \cite{M93} in computer
simulations, these instabilities can be well described by means of a linear stability
analysis of the Navier-Stokes hydrodynamic equations. An important feature of these instabilities is that they are confined to long wavelengths (small wave numbers) and hence, they can be suppressed by considering small enough systems. The linear stability analysis gives a critical length $L_c$ \cite{BDKS98,G05} such that the system becomes unstable when its linear size is larger than $L_c$. Given that the Navier-Stokes transport coefficients are involved in the evaluation of $L_c$, the determination of $L_c$ is perhaps one of the most nice applications of the Navier-Stokes hydrodynamics. In fact, the theoretical predictions of $L_c$ for monocomponent \cite{BDKS98,G05} and multicomponent \cite{BM13,G15} systems have been shown to agree in general very well with computer simulations for dilute \cite{BM13} and moderately dense \cite{MDCPH11,MGHEH12,MGH14} granular fluids.

On the other hand, although in nature granular matter is usually surrounded by an interstitial fluid (like the air, for instance), most theoretical and computational studies have ignored the effect of the latter on the dynamics properties of grains. However, under certain circumstances, the impact of the gas phase on the solid particles can be relevant in a wide range of practical applications, like for instance species segregation in granular mixtures (see for instance, refs.\ \cite{MLNJ01,NSK03,SSK04,WZXS08,CPSK10,PGM14}). At a kinetic theory level, the description of granular suspensions is an intricate problem since it involves two phases (solid particles and interstitial fluid) and hence, one would need
to solve a set of two coupled kinetic equations for each one of the velocity distribution functions of the different phases.
Thus, due to the mathematical difficulties embodied in this approach and in order to gain some insight into this problem,
an usual model \cite{KH01} for gas-solid flows is to consider a kinetic equation for the solid particles where
the influence of the surrounding fluid on them is modeled by means of an effective external force. This will be the approach
considered in the present paper.

An interesting problem is to assess the impact of the gas phase on the value of the critical length $L_c$ for the onset of instability. This problem has been addressed very recently in Ref.\ \cite{GFHY16}. These authors model the influence of the interstitial viscous gas on grains by means of a drag force $\mathbf{F}_{\text{fluid}}$ of the form  
\beq
\label{1.1}
\mathbf{F}_{\text{fluid}}=-m\gamma \left(\mathbf{v}-\mathbf{U}\right),
\eeq
where $m$ and $\mathbf{v}$ are the mass and the velocity of particles, respectively, $\gamma$ is the friction coefficient, and   $\mathbf{U}$ is the mean flow velocity [defined below in Eq.\ \eqref{2.6}]. The model defined by Eq.\ \eqref{1.1} (referred to as the thermal drag model) can be seen as a simplified version of the particle acceleration model proposed in Ref.\ \cite{GTSH12} where the effect of the gas phase is not only taken into account by the drag force \eqref{1.1} but also by means of two additional terms: (i) one term proportional to the difference $\Delta \mathbf{U}=\mathbf{U}-\mathbf{U}_g$ between the mean flow velocities of the gas ($\mathbf{U}_g$) and solid ($\mathbf{U}$) phases and (ii) a Langevin like-term (proportional to the parameter $\xi$) that accounts for added affects coming from neighboring particles. The parameter $\xi$ is proportional to the difference $\Delta \mathbf{U}$. Thus, when one considers particular situations (like the simple shear flow state \cite{TK95,SMTK96,ChVG15}) where the mean flow velocity of the solid particles follows the mean flow velocity of the gas phase, the difference $\Delta \mathbf{U}=\mathbf{0}$. In this case, the complete model proposed in Ref.\ \cite{GTSH12} reduces to the thermal drag model defined by Eq.\ \eqref{1.1}.

An important aspect of the model is the dependence of the friction coefficient $\gamma$ on the parameters of the system (particle mass and diameter, gas viscosity and hydrodynamic mean variables like solids concentration and granular temperature). In particular, in the case of low mean-flow Reynolds numbers, $\gamma$ is assumed to have a nonlinear dependence on the solid volume fraction $\phi$ but it is independent of the granular temperature $T$. This was the form of the friction coefficient adopted in the calculations performed in Ref.\ \cite{GFHY16}. A consequence of this choice is that the scaled friction coefficient $\gamma^*=\gamma/\nu$ (being $\nu \propto \sqrt{T}$ an effective collision frequency for hard spheres) depends on time through its dependence on the temperature $T$. This fact introduces a technical difficulty in the linear stability analysis of the hydrodynamic equations since the time dependence of the Navier-Stokes transport coefficients cannot be completely eliminated after changing space and time and scaling the hydrodynamic fields. This is essentially due to the different time scale of the drag coefficient $\gamma$ and the remaining time dependent parameters involved in the problem. As a consequence, the stability analysis made in Ref.\ \cite{GFHY16} required a numerical integration of the hydrodynamic equations.

A way to overcome the above technical difficulty is to assume that the friction coefficient is proportional to the collision frequency $\nu$ so that, the scaled parameter $\gamma^*\equiv \text{const}$. This new dependence opens the possibility of determining \emph{analytically} the critical length $L_c$ for gas-solid flows in terms of the solid volume fraction $\phi$, the coefficient of restitution $\al$ and the scaled friction coefficient $\gamma^*$. The simplification $\gamma^*\equiv \text{const}$ is essentially motivated by a desire for analytic expressions and the possibility of showing more clearly the impact of the gas phase on the value of $L_c$. Moreover, the good qualitative agreement found between the transport coefficients derived when $\gamma^*\equiv \text{const}$ and those obtained in Ref.\ \cite{GFHY16} encourages the use of the thermal drag model with $\gamma^*\equiv \text{const}$. The aim of this paper is to analyze instabilities in granular suspensions by considering the latter version of the model.   

The plan of the paper is as follows. In section \ref{sec2} the thermal drag model is introduced and the corresponding balance equations for the densities of mass, momentum, and energy are derived. Section \ref{sec3} deals with the application of the Chapman-Enskog method to solve the Enskog equation up to first order in the spatial gradients (Navier-Stokes hydrodynamic order). The explicit dependence of the transport coefficients on both inelasticity and viscous dissipation is illustrated for several systems. The results indicate that the influence of the gas phase on transport is in general significant. The linear stability analysis around the so-called homogeneous cooling state (HCS) of the hydrodynamic equations is worked out in section \ref{sec4}. This section presents  the main results of the paper. It is shown that in general the main effect of the gas phase on the critical length $L_c$ is to decrease its value with respect to the one obtained for dry granular fluids \cite{BDKS98,G05}. This conclusion agrees qualitatively well with computer simulations \cite{GFHY16}. In addition, as a complementary result, section \ref{sec5} displays the expressions of the transport coefficients when the drag force is employed as a thermostatic force to achieve a steady state. Finally, the paper is closed in Sec.\ \ref{sec6} with some concluding remarks.

\section{Enskog kinetic theory for gas-solid flows}
\label{sec2}

Let us consider a granular fluid modeled as a gas of smooth inelastic hard disks ($d=2$) or spheres ($d=3$). The inelasticity of collisions among all pairs is accounted for by a \emph{constant} (positive) coefficient of restitution $\al\leq 1$. As mentioned in the Introduction, a simple way of assessing the effect of the gas phase on the transport properties of grains is by means of the action of external nonconservative forces. These forces are incorporated into the corresponding kinetic equation of the solid particles. For moderate densities, the one-particle distribution function $f(\mathbf{r},\mathbf{v},t)$ of grains obeys the Enskog kinetic equation \cite{BDS97,BP04}
\begin{equation}
\label{2.1}
\frac{\partial f}{\partial t}+{\bf v}\cdot \nabla f+\frac{\partial}{\partial \mathbf{v}}\cdot \left(\frac{\mathbf{F}_{\text{fluid}}}{m}f\right)=J_\text{E}[\mathbf{r},\mathbf{v}|f,f],
\end{equation}
where
\beqa
\label{2.2}
J_{\text{E}}\left[{\bf r}, {\bf v}_{1}|f,f\right] &=&\sigma^{d-1}\int \dd{\bf v}
_{2}\int \dd\widehat{\boldsymbol{\sigma}}\,\Theta (\widehat{{\boldsymbol {\sigma }}}
\cdot {\bf g}_{12})(\widehat{\boldsymbol {\sigma }}\cdot {\bf g}_{12})\left[ \alpha^{-2}
\chi({\bf r},{\bf r}-\boldsymbol {\sigma })
f({\bf r}, {\bf v}_1';t)\right. \nonumber\\
& & \times \left.f({\bf r}-\boldsymbol {\sigma}, {\bf v}_2';t)-
\chi({\bf r},{\bf r}+\boldsymbol {\sigma })
f({\bf r}, {\bf v}_1;t) f({\bf r}+\boldsymbol {\sigma }, {\bf v}_2;t)\right]
\eeqa
is the Enskog collision operator. Like the Boltzmann equation, the Enskog equation neglects velocity correlations among particles that are about to collide, but it takes into account the dominant spatial correlations due to excluded-volume effects. Here, $d$ is the dimensionality of the system ($d=2$ for disks and $d=3$ for spheres), $\sigma$ is the hard sphere diameter,  $\boldsymbol {\sigma}=\sigma \widehat{\boldsymbol {\sigma}}$,  $\widehat{\boldsymbol {\sigma}}$ being a unit vector along the line of centers, $\Theta$ is the Heaviside step function, and ${\bf g}_{12}={\bf v}_{1}-{\bf v}_{2}$ is the relative velocity of the two colliding spheres. Moreover, $\chi[{\bf r},{\bf r}+\boldsymbol {\sigma}|\{n(t)] $ is the equilibrium pair correlation function at contact as a functional of the nonequilibrium density field $n({\bf r}, t)$ defined by
\beq
\label{2.3}
n({\bf r}, t)=\int\; \dd{\bf v} f({\bf r},{\bf v},t).
\eeq
The primes on the velocities in Eq.\ \eqref{2.1} denote the initial values $\{{\bf v}_{1}',
{\bf v}_{2}'\}$ that lead to $\{{\bf v}_{1},{\bf v}_{2}\}$
following a binary collision:
\begin{subequations}
\begin{equation}
{\bf v}_{1}'={\bf v}_{1}-\frac{1}{2}\left( 1+\alpha^{-1}\right)
(\widehat{{\boldsymbol {\sigma }}}\cdot {\bf g}_{12})\widehat{{\boldsymbol {\sigma }}},
\eeq
\beq
{\bf v}_{2}'={\bf v}_{2}+\frac{1}{2}\left( 1+\alpha^{-1}\right)
(\widehat{{\boldsymbol {\sigma }}}\cdot {\bf g}_{12})\widehat{
\boldsymbol {\sigma}}. \label{2.4}
\end{equation}
\end{subequations}

A simplest way of modeling the fluid-solid force $\mathbf{F}_{\text{fluid}}$ in high-velocity gas-solid flows is through a viscous drag force of the form \eqref{1.1} where
\begin{equation}
\label{2.6}
\mathbf{U}({\bf r}, t)=\frac{1}{n({\bf r}, t)}\int \;\dd{\bf v} \; {\bf v}\;  f(\mathbf{r},{\bf v},t)
\end{equation}
is the mean flow velocity of solid particles. The friction coefficient $\gamma$ is proportional to the viscosity of the surrounding gas $\mu_g$ and is assumed to be a known quantity. Thus, according to Eqs.\ \eqref{1.1} and \eqref{2.1}, the Enskog equation for the thermal drag model reads
\beq
\label{2.8}
\frac{\partial f}{\partial t}+{\bf v}\cdot \nabla f-\gamma \frac{\partial}{\partial \mathbf{V}}\cdot \mathbf{V}f=J_\text{E}[\mathbf{r},\mathbf{v}|f,f],
\eeq
where $\mathbf{V}=\mathbf{v}-\mathbf{U}$ is the peculiar velocity. As said in the Introduction, the thermal drag model \eqref{2.8} can be seen as a simplified version of the model proposed in Ref.\ \cite{GTSH12} since the former reduces to the latter when the mean flow velocities of solid particles ($\mathbf{U}$) and gas phase ($\mathbf{U}_g$) coincide. In this case ($\Delta \mathbf{U}\equiv \mathbf{U}-\mathbf{U}_g=\mathbf{0}$), the mean drag (which is proportional to $\Delta \mathbf{U}$) and neighbor effects (which are accounted for by a Langevin-like term) are assumed to be negligible in the model proposed in Ref.\ \cite{GTSH12}. In this context, the results derived here can be considered as relevant for transport problems where the difference between the mean flow velocities of solid and gas phases is very small. It is worthwhile remarking that the thermal drag model \eqref{2.8} has been previously employed to analyze simple shear flows in gas-solid suspensions \cite{TK95,SMTK96,ChVG15}, particle clustering due to hydrodynamic interactions \cite{WK00},  steady states driven by vibrating walls \cite{WZLH09}, and more recently \cite{H13,SMMD13,WGZS14} to study rheology of frictional hard-sphere suspensions.

Another relevant conceptual point is that the form of the collision operator $J_\text{E}[\mathbf{r},\mathbf{v}|f,f]$ in the suspension model \eqref{2.8} is the same as for a \emph{dry} granular fluid and so, the gas phase does not play any role in the collision dynamics. As has been previously discussed in several papers \cite{K90,TK95,SMTK96,WKL03}, the
above assumption requires that the mean-free time between collisions is much less than the viscous relaxation time due to fluid forces. Under these conditions, it is expected that the suspension model defined by Eq.\ \eqref{2.8} will accurately describe situations where the stresses exerted by the gas phase have a small influence on the dynamics of solid particles. 

The other relevant hydrodynamic velocity moment of the distribution $f$ is the so-called \emph{granular} temperature. It is defined as
\begin{equation}
\label{2.9}
T({\bf r}, t)=\frac{m}{d n({\bf r}, t)}\int \; \dd{\bf v}\; V^2\; f(\mathbf{r},{\bf v},t).
\end{equation}
The macroscopic balance equations for the densities of mass, momentum and energy can be exactly derived from the Enskog equation \eqref{2.1}. The are given by \cite{GTSH12}
\begin{equation}
D_{t}n+n\nabla \cdot {\bf U}=0\;, \label{2.10}
\end{equation}
\begin{equation}
D_{t}{\bf U}+\rho ^{-1}\nabla \cdot {\sf P}=\mathbf{0}\;,
\label{2.11}
\end{equation}
\begin{equation}
D_{t}T+\frac{2}{dn} \left( \nabla \cdot {\bf q}+{\sf P}:\nabla {\bf U}\right) =
-\left(2\gamma +\zeta\right) \,T\;.
\label{2.12}
\end{equation}
In the above equations, $D_{t}=\partial_{t}+{\bf U}\cdot \nabla$ is the material
derivative and $\rho = m n$ is the mass density. The presence of the gas phase gives rise to the thermal drag term [first term on the right hand side of Eq.\ \eqref{2.12}] which accounts for the energy dissipated per unit time due to the surrounding gas viscosity. In addition, the cooling rate $\zeta$ is proportional to $1-\alpha^2$ and is due to dissipative collisions. The pressure or stress tensor ${\sf P}({\bf r},t)$ and the heat flux ${\bf q}({\bf r},t)$ have both {\em kinetic} and {\em collisional transfer} contributions, i.e., ${\sf P}={\sf P}^k+{\sf P}^c$ and ${\bf q}={\bf q}^k+{\bf q}^c$. The kinetic contributions are given by
\begin{equation}
\label{2.13}
{\sf P}^k=\int \; \dd{\bf v} m{\bf V}{\bf V}f({\bf r},{\bf v},t),
\end{equation}
\begin{equation}
\label{2.14}
{\bf q}^k=\int \; \dd{\bf v} \frac{m}{2}V^2{\bf V}f({\bf r},{\bf v},t).
\end{equation}
The definition of the collisional transfer contributions ${\sf P}^c$ and $\mathbf{q}^c$ are given by Eqs.\ (4.11) and (4.12), respectively, of Ref.\ \cite{GTSH12}. Since the forms of the collisional contributions to ${\sf P}$ and $\mathbf{q}$ in the first order in spatial gradients (Navier-Stokes hydrodynamic order) are the same as those previously derived in Ref.\ \cite{GTSH12}, we do not include them here for the sake of brevity. The cooling rate is given by
\beq
\zeta=\frac{\left(1-\alpha^{2}\right)}{4dnT} m \sigma^{d-1}\int \dd\mathbf{v}
_{1}\int \dd\mathbf{v}_{2}\int \dd\widehat{\boldsymbol {\sigma }}
\Theta (\widehat{\boldsymbol {\sigma }}\cdot
\mathbf{g}_{12}) (\widehat{ \boldsymbol {\sigma }}\cdot
\mathbf{g}_{12})^{3}\chi(\mathbf{r},\mathbf{r}+\boldsymbol {\sigma})f(\mathbf{r},\mathbf{v},t)f(\mathbf{r}+\boldsymbol {\sigma},\mathbf{v},t).
\label{2.15}
\eeq

In order to close the hydrodynamic balance equations \eqref{2.10}--\eqref{2.12}, the constitutive equations for the stress tensor, the heat flux, and the cooling rate must be known. This means that $\mathsf{P}$, $\mathbf{q}$, and $\zeta$ must be expressed as functionals of the hydrodynamic fields $n$, $\mathbf{U}$, and $T$. This task can be accomplished by solving the Enskog equation \eqref{2.8} by means of the Chapman-Enskog method \cite{CC70} conveniently adapted to account for the dissipative dynamics. In particular, in the first-order of the expansion, the stress tensor $P_{ij}^{(1)}$ and the heat flux $\mathbf{q}^{(1)}$ are given, respectively, as
\begin{equation}
\label{2.16}
P_{ij}^{(1)}=-\eta\left( \partial_{i}U_{j}+\partial _{j
}U_{i}-\frac{2}{d}\delta _{ij}\nabla \cdot
\mathbf{U} \right) -\eta_b  \nabla \cdot
\mathbf{U},
\end{equation}
\beq
\label{2.17}
\mathbf{q}^{(1)}=-\kappa \nabla T-\mu \nabla n.
\eeq
Here, $\eta$ is the shear viscosity, $\eta_b$ is the bulk viscosity, $\kappa$ is the thermal conductivity coefficient, and $\mu$ is a Dufour-like coefficient. While $\eta$, $\kappa$ and $\mu$ have kinetic and collisional contributions, $\eta_b$ has only a collisional contribution. In particular, the kinetic contributions to the Navier-Stokes transport coefficients can be written in the form
\beq
\label{2.18}
\eta_k(T)=\eta_0(T) \eta_k^*, \quad \kappa_k(T) = \kappa_0(T) \kappa_k^*, \quad \mu_k(T)= \frac{T\kappa_0(T)}{n}\mu_k^*,
\eeq
where $\eta_0(T)$ and $\kappa_0(T)$ are the shear viscosity and the thermal conductivity coefficients, respectively, of a molecular (dry) dilute gas. Their expressions are
\beq
\label{2.19}
\eta_0=\frac{n T}{\nu}, \quad \kappa_0=\frac{d(d+2)}{2(d-1)}\frac{\eta_0}{m},
\eeq
where
\beq
\label{2.20}
\nu(T)=\frac{8}{d+2}\frac{\pi^{\left(d-1\right) /2}}{\Gamma \left( \frac{d}{2}\right)}
n\sigma^{d-1}\sqrt{\frac{T}{m}}
\eeq
is the collision frequency associated with the shear viscosity of a dilute elastic gas.

The (scaled) transport coefficients $\eta_k^*$, $\kappa_k^*$, and $\mu_k^*$ are nonlinear functions of the solid volume fraction
\beq
\label{2.20.1}
\phi=\frac{\pi ^{d/2}}{2^{d-1}d\Gamma \left(\frac{d}{2}\right)}n\sigma^d,
\eeq
the coefficient of restitution $\al$, and the (dimensionless) friction coefficient $\gamma^* \equiv \gamma/\nu$. In the case that $\gamma \equiv \text{const.}$, the scaled coefficient $\gamma^*$ depends on time through its dependence on the granular temperature [$\gamma^*(t) \propto T(t)^{-1/2}$]. In this case, in contrast to the results obtained for a dry granular fluid \cite{GD99a,L05}, the scaled transport coefficients have an additional time dependence through the temperature dependence of $\gamma^*$. The explicit forms of $\eta^*$, $\lambda^*$, $\kappa^*$, and $\mu^*$ by considering $\gamma \equiv \text{const.}$ were derived in Ref.\ \cite{GFHY16}. In addition, apart from evaluating transport, a linear stability analysis of the hydrodynamic equations was also performed in Ref.\ \cite{GFHY16} to analyze the stability of the HCS. On the other hand, as mentioned in the Introduction and in contrast to previous stability analysis \cite{BDKS98,G05,GMD06} for (dry) granular gases, the time dependence of the scaled transport coefficients could not completely be eliminated after changing the time and space variables and scaling the hydrodynamic fields. As a consequence, only numerical results for the transversal shear mode (and not for the remaining longitudinal modes) were reported in the above work \cite{GFHY16}. Thus, in order to make some analytical progress on the stability analysis of the HCS, one could assume that the friction coefficient $\gamma$ is not constant and is proportional to the effective collision frequency $\nu \propto n \sqrt{T}$. In this case, $\gamma^*\equiv \text{const.}$ This choice allows us to offer a complete analytical study of instabilities in granular gas-solid flows.

\section{Navier-Stokes transport coefficients}
\label{sec3}

To determine the Navier-Stokes transport coefficients, let us assume that the HCS is slightly perturbed by small spatial gradients. These perturbations give rise to nonzero contributions to the momentum and heat fluxes and hence, one can identify the expressions of the relevant transport coefficients as functions of both the coefficient of restitution $\al$ and the friction coefficient $\gamma$. Since the strength of the spatial gradients is small, the Enskog kinetic equation \eqref{2.8} can be solved by the Chapman-Enskog expansion \cite{CC70} up to first order in spatial gradients. The Chapman-Enskog method assumes the existence of a normal or hydrodynamic solution where all space and time dependence of the distribution $f$ only occurs through the hydrodynamic fields. This functional dependence can be made local in space by means of an expansion in gradients of the hydrodynamic fields $n$, $\mathbf{U}$, and $T$. Thus, $f$ can be written as
\beq
\label{3.1}
f=f^{(0)}+f^{(1)}+\ldots,
\eeq
where the approximation $f^{(k)}$ is of order $k$ in spatial gradients. Moreover, collisional dissipation is assumed to be uncoupled of the spatial gradients and so, we consider situations where the spatial gradients are sufficiently small (low Knudsen number) but there is no \emph{a priori} any limitation on the degree of collisional dissipation. In addition, in ordering the different level of approximations in the kinetic equation, as in previous papers \cite{GTSH12,GFHY16}, the friction coefficient $\gamma$ will be taken to be of zeroth order in gradients since it does not induce any flux in the system.

The zeroth-order solution $f^{(0)}$ is the local version of the HCS distribution. It can be written as \cite{GFHY16}
\beq
\label{3.2}
f^{(0)}(\mathbf{r},\mathbf{v},t)=n(\mathbf{r},t)v_0(\mathbf{r},t)^{-d} \varphi(\mathbf{c}),
\eeq
where $v_0=\sqrt{2T/m}$ is the thermal velocity and $\mathbf{c}\equiv \mathbf{V}/v_0$. Note that even when $\gamma\equiv \text{const}$, according to the previous results derived for driven granular gases \cite{GMT12,GChV13,ChVG13}, one could expect that the scaled distribution $\varphi$ could have an additional dependence of the granular temperature through the dimensionless friction coefficient $\gamma/\nu(T)$. However, it can easily seen by direct substitution that the form \eqref{3.2} is still a solution of the homogeneous version of the Enskog equation \eqref{2.8}. This conclusion agrees with the results obtained in Ref.\ \cite{L01} where the drag force term $\partial_\mathbf{v} \cdot \mathbf{v} f$ was shown to arise from a logarithmic change in the time scale of the hard-sphere system in the absence of the drag force.

\begin{figure}
{\includegraphics[width=0.4\columnwidth]{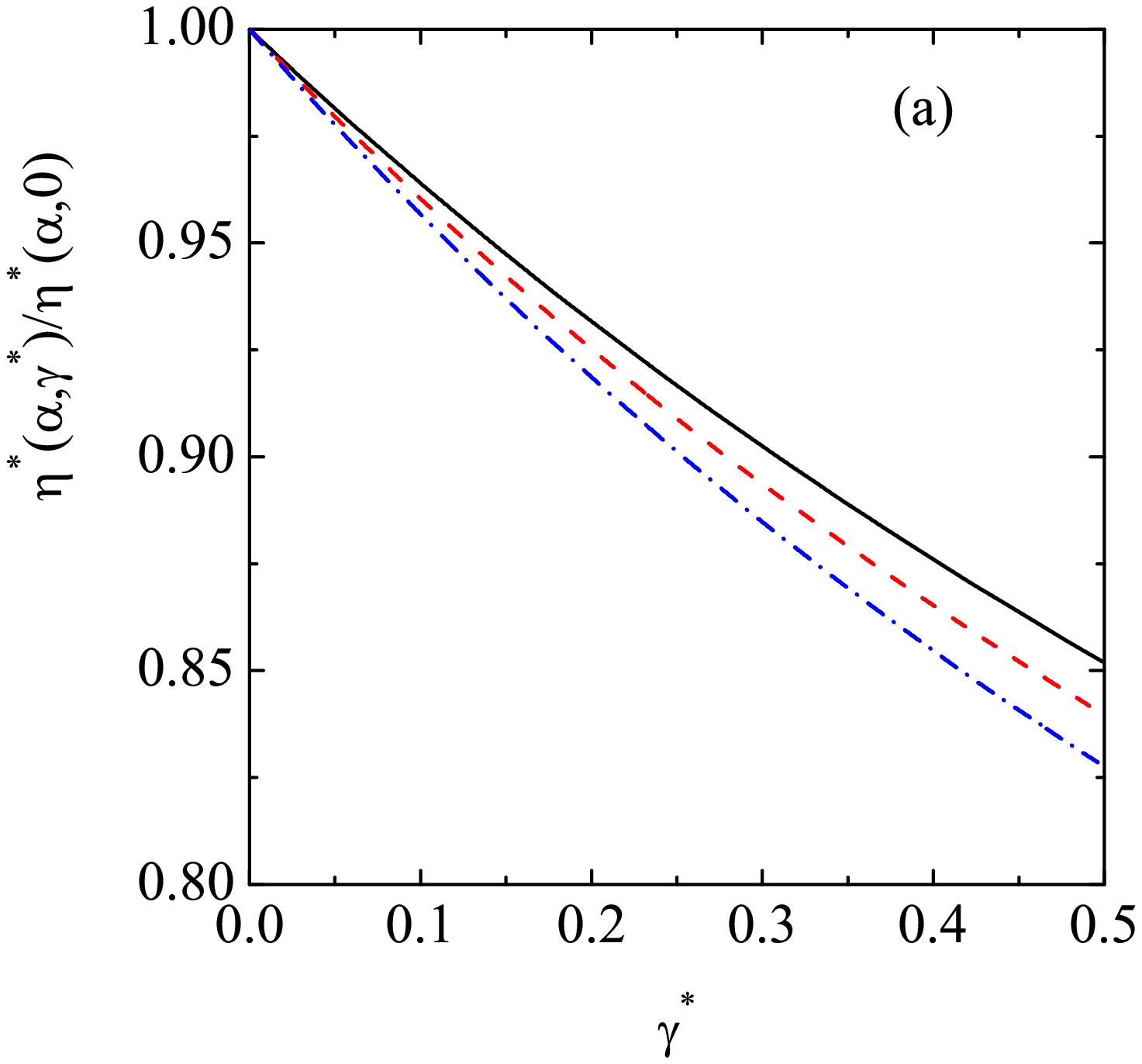}}
{\includegraphics[width=0.4\columnwidth]{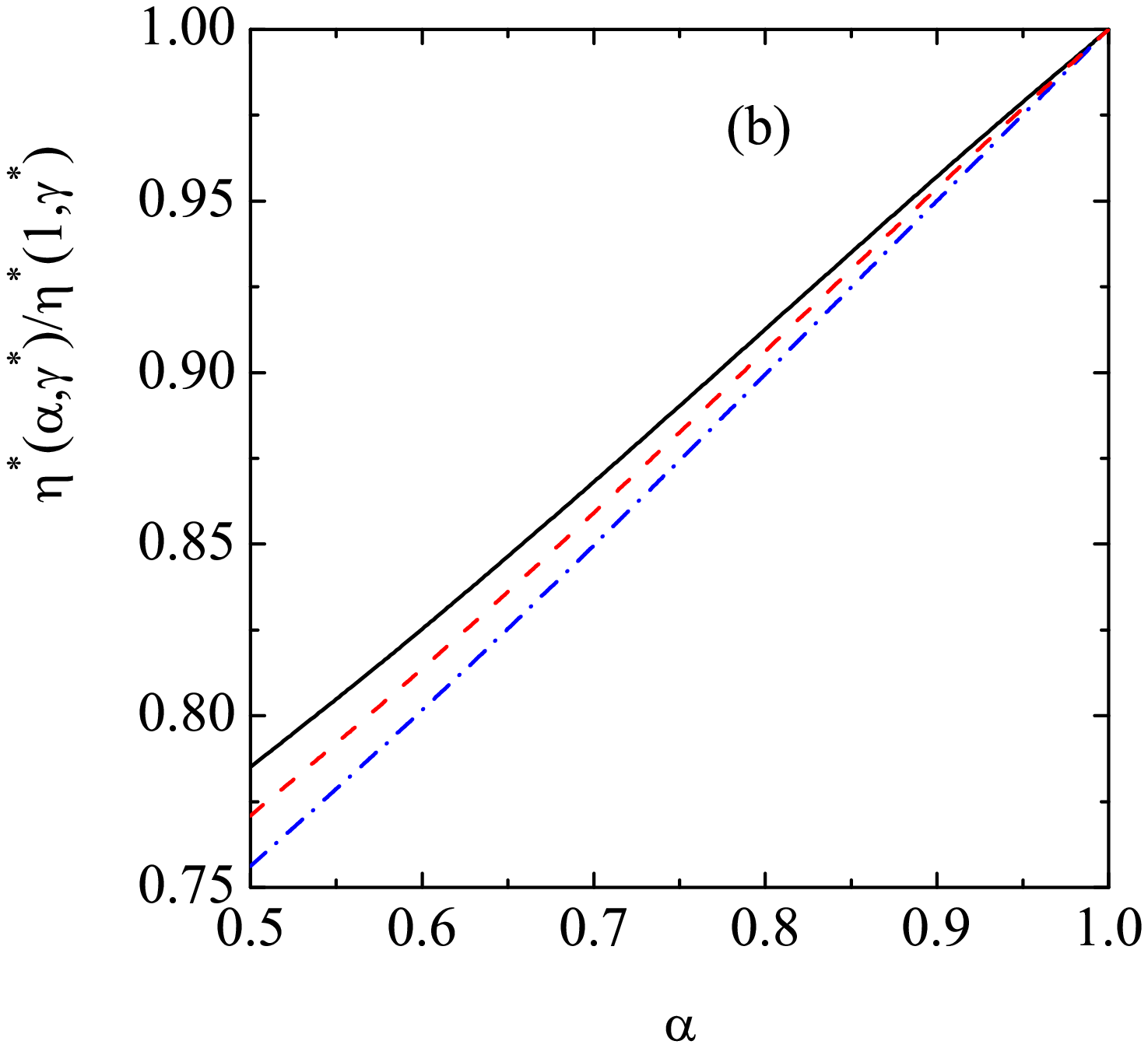}}
\caption{(Color online) (a) Plot of the ratio $\eta^*(\al,\gamma^*)/\eta^*(\alpha,0)$ versus the (dimensionless) friction coefficient $\gamma^*$ for $d=3$, $\phi=0.2$ and three different values of the coefficient of restitution $\alpha$: $\alpha=1$ ( solid line), $\alpha=0.8$ (dashed line), and $\alpha=0.6$ (dash-dotted line). (b) Plot of the ratio $\eta^*(\al,\gamma^*)/\eta^*(1,\gamma^*)$ versus the coefficient of restitution $\alpha$ for $d=3$, $\phi=0.2$ and three different values of the (dimensionless) friction coefficient $\gamma^*$: $\gamma^*=0$ (solid line), $\gamma^*=0.2$ (dashed line), and $\gamma^*=0.5$ (dash-dotted line).
\label{eta}}
\end{figure}
\begin{figure}
{\includegraphics[width=0.4\columnwidth]{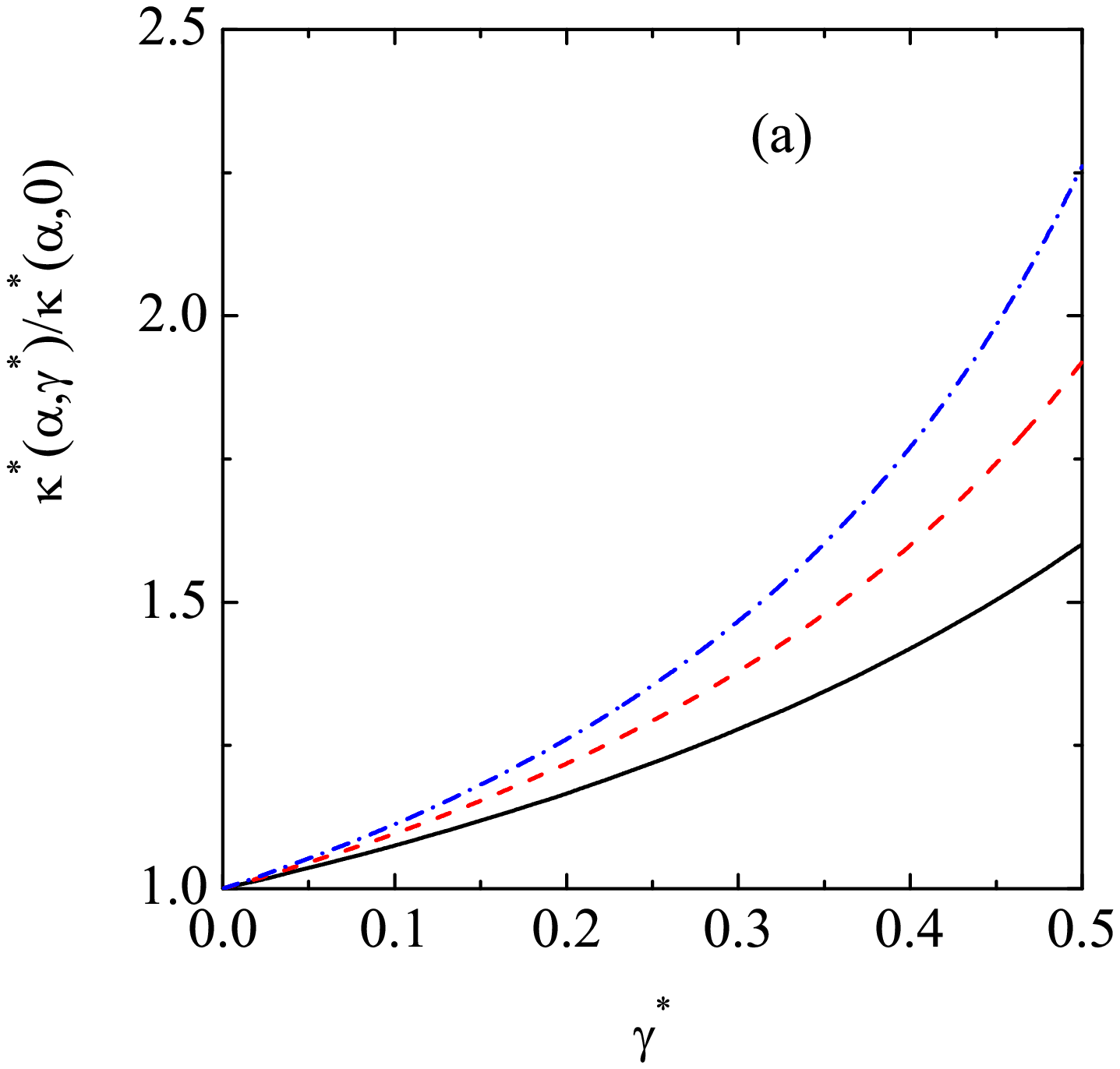}}
{\includegraphics[width=0.4\columnwidth]{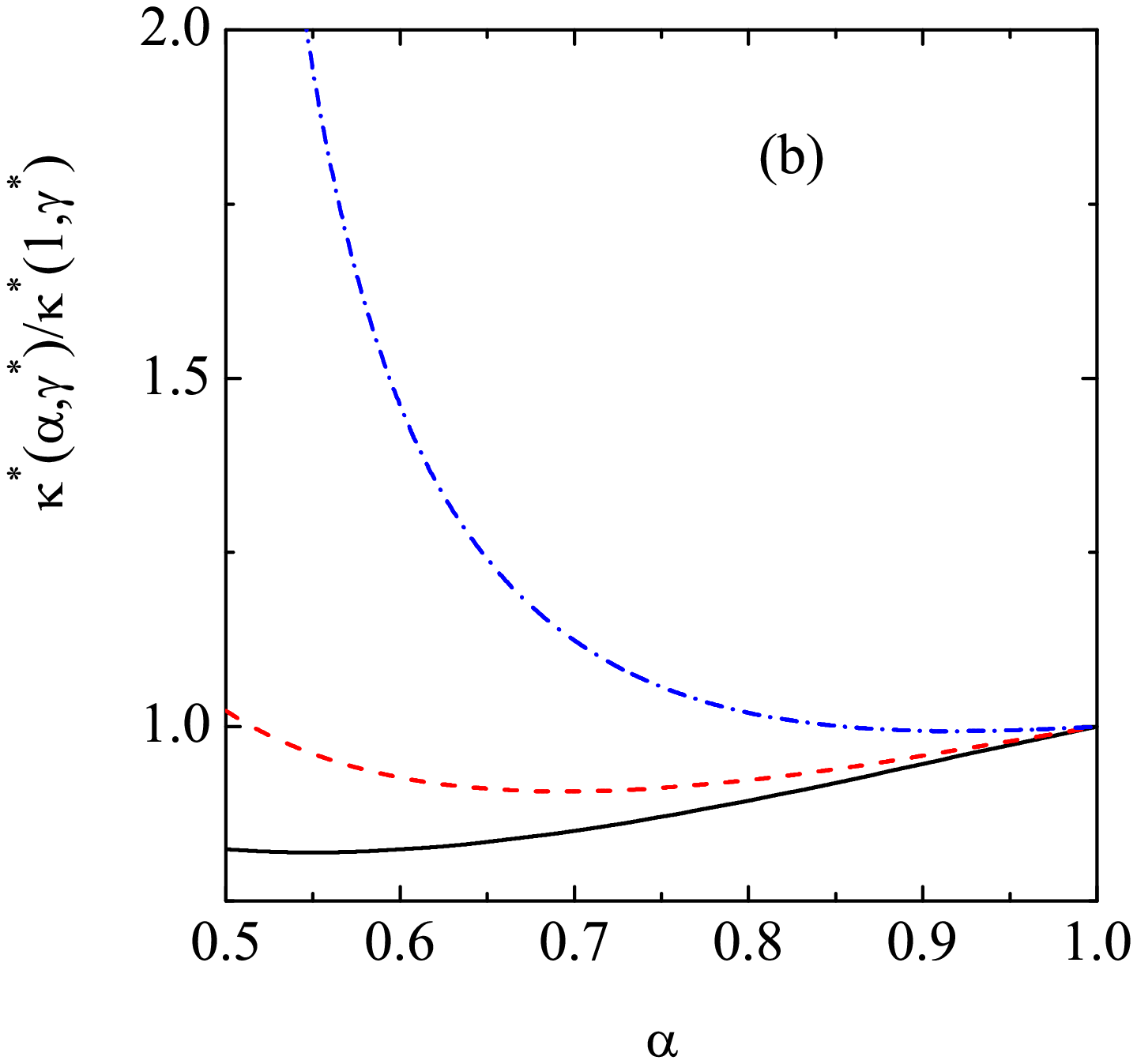}}
\caption{(Color online) (a) Plot of the ratio $\kappa^*(\al,\gamma^*)/\kappa^*(\alpha,0)$ versus the (dimensionless) friction coefficient $\gamma^*$ for $d=3$, $\phi=0.2$ and three different values of the coefficient of restitution $\alpha$: $\alpha=1$ ( solid line), $\alpha=0.8$ (dashed line), and $\alpha=0.6$ (dash-dotted line). (b) Plot of the ratio $\kappa^*(\al,\gamma^*)/\kappa^*(1,\gamma^*)$ versus the coefficient of restitution $\alpha$ for $d=3$, $\phi=0.2$ and three different values of the (dimensionless) friction coefficient $\gamma^*$: $\gamma^*=0$ (solid line), $\gamma^*=0.2$ (dashed line), and $\gamma^*=0.5$ (dash-dotted line).
\label{kappa}}
\end{figure}
\begin{figure}
{\includegraphics[width=0.4\columnwidth]{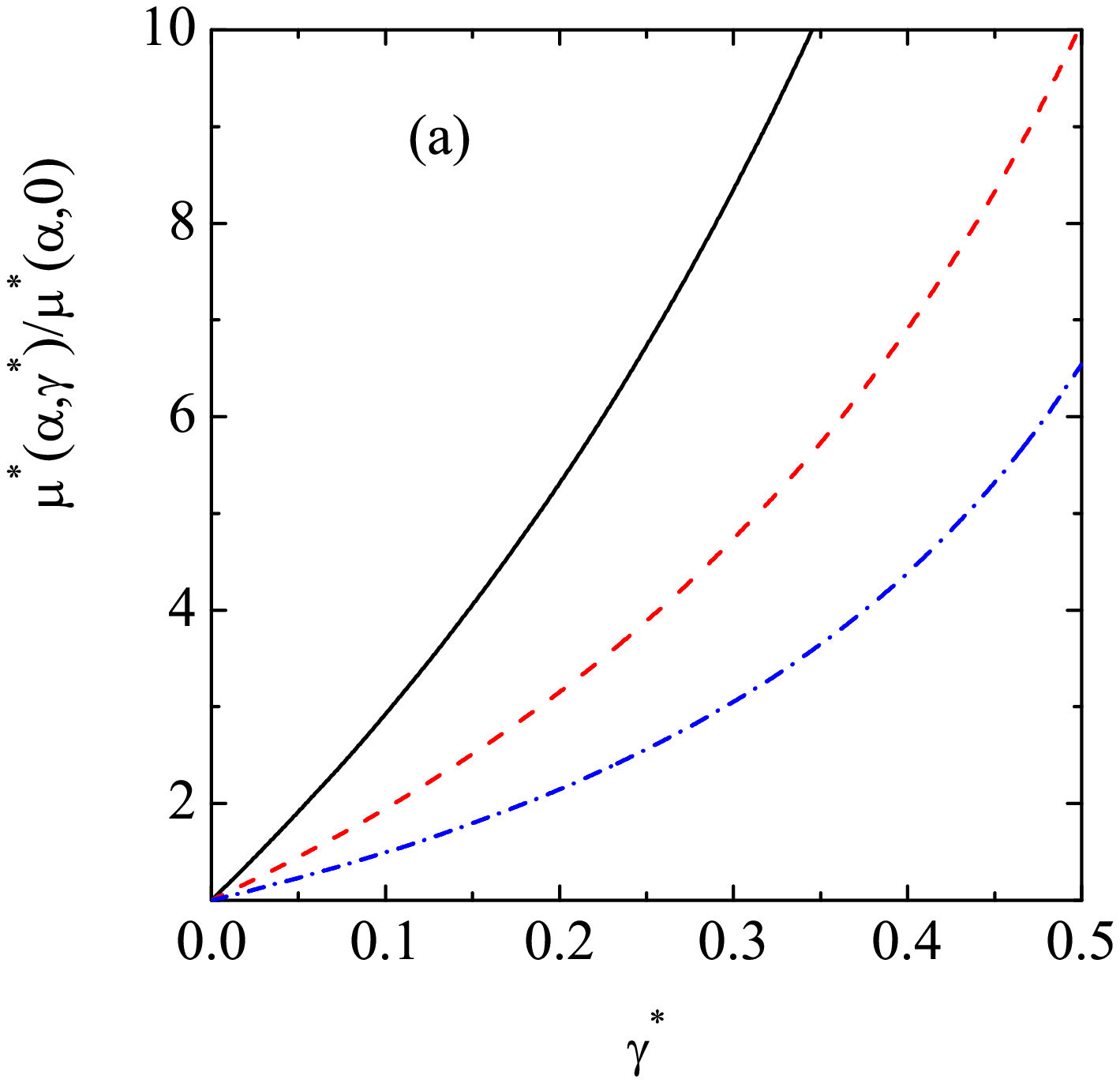}}
{\includegraphics[width=0.4\columnwidth]{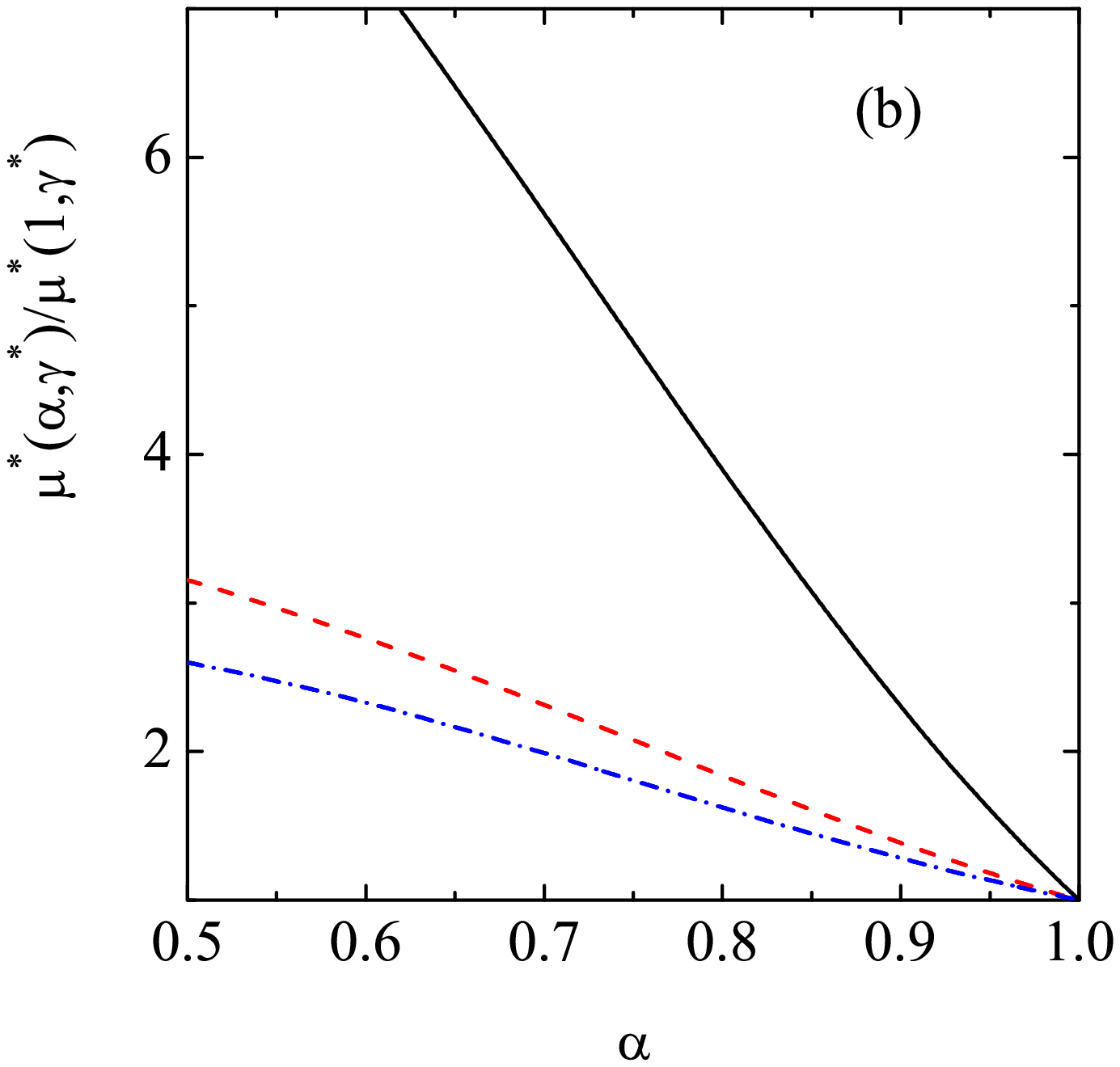}}
\caption{(Color online) (a) Plot of the ratio $\mu^*(\al,\gamma^*)/\mu^*(\alpha,0)$ versus the (dimensionless) friction coefficient $\gamma^*$ for $d=3$, $\phi=0.2$ and three different values of the coefficient of restitution $\alpha$: $\alpha=0.9$ (solid line), $\alpha=0.8$ (dashed line), and $\alpha=0.6$ (dash-dotted line). (b) Plot of the ratio $\eta^*(\al,\gamma^*)/\eta^*(1,\gamma^*)$ versus the coefficient of restitution $\alpha$ for $d=3$, $\phi=0.2$ and three different values of the (dimensionless) friction coefficient $\gamma^*$: $\gamma^*=0.05$ (solid line), $\gamma^*=0.2$ (dashed line), and $\gamma^*=0.5$ (dash-dotted line). Note that for a dry molecular fluid ($\al=1$ and $\gamma=0$) the coefficient $\mu$ vanishes ($\mu^*(1,0)=0$).
\label{mu}}
\end{figure}

The analysis to first order in the Chapman-Enskog method follows similar steps as those performed before in several papers \cite{GD99a,GTSH12,GFHY16}. Only the final forms of the Navier-Stokes transport coefficients and the cooling rate are displayed here, with some technical details being given in Appendix \ref{appA}. The expressions of the (scaled) transport coefficients $\eta^*\equiv \eta/\eta_0$, $\eta_b^*\equiv \eta_b/\eta_0$, $\kappa^*\equiv \kappa/\kappa_0$, and $\mu^* \equiv (\kappa_0 T/n)\mu$ are
\begin{equation}
\label{3.3}
\eta^*=\eta_k^*\left[1+\frac{2^{d-1}}{d+2}\phi \chi \left(1+\alpha\right)\right]+\frac{d}{d+2}\eta_b^*,
\end{equation}
\begin{equation}
\label{3.4}
\eta_b^*=\frac{2^{2d+1}}{\pi(d+2)}\phi^2 \chi (1+\alpha)\left(1-\frac{a_2}{16} \right),
\end{equation}
\beq
\label{3.5}
\kappa^*=\kappa_{k}^*\left[1+3\frac{2^{d-2}}{d+2}\phi \chi (1+\alpha)\right]+\frac{2^{2d+1}(d-1)}{(d+2)^2\pi}
\phi^2 \chi (1+\alpha)\left(1+\frac{7}{16} a_2 \right),
\eeq
\begin{equation}
\label{3.6}
\mu^*=\mu_k^*\left[1+3\frac{2^{d-2}}{d+2}\phi \chi (1+\alpha)\right],
\end{equation}
where the kinetic coefficients $\eta_k^*$, $\kappa_k^*$, and $\mu_k^*$ are given by Eqs.\ \eqref{a12}--\eqref{a14}, respectively, and
\beq
\label{3.18}
a_{2}=\frac{16(1-\al)(1-2\al^2)}{9+24d-\al(41-8d)+30\al^2(1-\al)}
\eeq
is the fourth cumulant of the zeroth-order distribution $f^{(0)}$ \cite{NE98}. Finally, to first order in gradients, the (reduced) cooling rate $\zeta^*\equiv \zeta/\nu_0$ is
\beq
\label{3.23}
\zeta^*=\zeta_0^*+\zeta_U \nabla \cdot \mathbf{U},
\eeq
where
\beq
\label{a16}
\zeta_0^*=\frac{d+2}{4d}\chi (1-\al^2)\left(1+\frac{3}{16}a_2\right),
\eeq
and a good approximation for $\zeta_U$ is \cite{GS95}
\beq
\label{3.24}
\zeta_U=-3\frac{2^{d-2}}{d}\phi \chi (1-\alpha^2).
\end{equation}
As for molecular fluids \cite{CC70,FK72}, the bulk viscosity $\eta_b^*$ presents a quadratic dependence on the solid volume fraction $\phi$ ($\eta_b^* \propto \phi^2$). This dependence is similar to the one found for the collisional contribution to the thermal conductivity $\kappa^*$ [see the second term on the right-hand side of Eq.\ \eqref{3.5}]. The quadratic dependence of $\eta_b^*$ on $\phi$ is essentially due to the fact that $\eta_b^*$ has \emph{only} collisional contributions and hence, its density dependence is stronger than the one obtained for the kinetic contributions $\eta_k^*$, $\kappa_k^*$, and $\mu_k^*$.

Figures \ref{eta}(a), \ref{kappa}(a), and \ref{mu}(a) show the dependence of the ratios $\eta^*(\al,\gamma^*)/\eta^*(\al,0)$, $\kappa^*(\al,\gamma^*)/\kappa^*(\al,0)$, and $\mu^*(\al,\gamma^*)/\mu^*(\al,0)$, respectively, on the (scaled) friction coefficient $\gamma^*$ for several values of the coefficient of restitution $\al$. As a complement, the ratios $\eta^*(\al,\gamma^*)/\eta^*(1,\gamma^*)$, $\kappa^*(\al,\gamma^*)/\kappa^*(1,\gamma^*)$, and $\mu^*(\al,\gamma^*)/\mu^*(1,\gamma^*)$ are plotted in Figs.\ \ref{eta}(b), \ref{kappa}(b), and \ref{mu}(b), respectively, as functions of $\al$ for several values of $\gamma^*$. With respect to the dependence of the (scaled) transport coefficients on $\gamma^*$, it appears that the impact of viscous gas on transport can be significant, specially in the case of the heat flux transport coefficients. Regarding the influence of collisional dissipation on transport, as expected from the results derived for dry granular gases \cite{GD99a}, we observe that the impact of $\al$ on the transport coefficients is in general important since their functional form differs appreciably from their elastic form. Moreover, Figs.\ \ref{eta}--\ref{mu} also indicate that while the results derived here for $\eta^*$ and $\mu^*$ agree qualitatively well with those obtained in Ref.\ \cite{GFHY16} by accounting for the temperature dependence of $\gamma^*$, more discrepancies appear for the thermal conductivity $\kappa^*$ since the latter was found to be independent of $\gamma^*$ in Ref.\ \cite{GFHY16}.

When the expressions of the pressure tensor, the heat flux and the cooling rate are substituted into the balance equations \eqref{2.10}--\eqref{2.12} one achieves the corresponding Navier-Stokes (closed) hydrodynamic equations for $n$, $\mathbf{U}$, and $T$. They are given by
\begin{equation}
\label{3.25}
D_tn+n\nabla \cdot {\bf U}=0,
\end{equation}
\beq
\label{3.26}
D_t U_i+\rho^{-1}\nabla_i p=\rho^{-1}\nabla_j\left[\eta \left(\nabla_i U_j+\nabla_j U_i-\frac{2}{d}
\delta_{ij}\nabla \cdot {\bf U}\right)+\eta_b \delta_{ij}\nabla \cdot {\bf U}\right],
\eeq
\beqa
\label{3.27}
n\left(D_t+2\gamma+\zeta^{(0)}\right)T+\frac{2}{d}p\nabla \cdot {\bf U}&=&\frac{2}{d}\nabla \cdot \left
(\kappa \nabla T+\mu \nabla n\right) +\frac{2}{d}\left[\eta \left(\nabla_i U_j+\nabla_j U_i-\frac{2}{d}
\delta_{ij}\nabla \cdot {\bf U}\right)\right.\nonumber\\
& & \left.
+\eta_b \delta_{ij}\nabla \cdot {\bf U}\right]\nabla_i U_j
-nT\zeta_U  \nabla\cdot {\bf U}.
\eeqa
Note that consistency would require to consider up to second order in the gradients in the expression \eqref{3.23}
for the cooling rate, since this is the order of the terms in Eq.\ \eqref{3.27} coming from
the pressure tensor and the heat flux. However, it has been shown for a granular dilute gas that the contributions from the cooling rate of second order are negligible as compared with the corresponding contributions from Eqs.\ \eqref{2.16}, \eqref{2.17}, and \eqref{3.23} \cite{BDKS98}. A similar behavior is expected in the case of suspensions at moderate densities.

The form of the Navier-Stokes equations \eqref{3.25}--\eqref{3.27} is the same as for a \emph{dry} granular fluid ($\gamma=0$), except for the presence of the friction coefficient $\gamma$ on the left hand side of the energy balance equation \eqref{3.27} and the dependence of the transport coefficients $\eta$, $\kappa$, and $\mu$ on $\gamma$.

\section{Linear stability analysis}
\label{sec4}

The hydrodynamic equations \eqref{3.25}--\eqref{3.27} admit a simple solution corresponding to the so-called HCS. It describes a uniform state with vanishing (or constant) flow field and a time-dependent temperature:
\beq
\label{4.1}
\frac{\partial T}{\partial t}=-\left(\zeta^{(0)}+2\gamma\right)T.
\eeq
Since $\gamma \propto \nu$, then the solution to Eq.\ \eqref{4.1} is
\beq
\label{4.2}
T(t)=\frac{T(0)}{\left[1+\frac{\zeta^{(0)}(0)+2\gamma(0)}{2}t\right]^2},
\eeq
where $T(0)$ is the initial temperature and $\zeta^{(0)}(0)$ and $\gamma(0)$ are the initial values of $\zeta^{(0)}$ and $\gamma$, respectively. On the other hand, as some computer simulations have previously shown for granular \cite{GZ93,M93,MY94,MY96,BRC99b,MDCPH11,MGHEH12,MGH14} and gas-solid \cite{WK00} flows, the HCS is unstable with respect to long  enough wavelength perturbations. This problem can be studied from a linear stability analysis of the hydrodynamic equations with respect to the HCS. This is the main objective of the present contribution.

To perform a linear stability analysis of Eqs.\ \eqref{3.25}--\eqref{3.27}, first one has to linearize those equations with respect to the time-dependent HCS state. As expected, this yields partial differential equations with coefficients that are independent of space but depend on time since the reference state is cooling. In the same way as previous stability analysis for dry granular fluids \cite{BDKS98,G05,BM13,G15,GMD06}, this time dependence can be eliminated through a change in the time and space variables and a scaling of the hydrodynamic fields. This is the main advantage of the analysis made here with respect to the previous one \cite{GFHY16} where the time dependence of $\gamma^*$ was not completely eliminated and only a numerical integration allowed to identify the critical size of the system for vortex instabilities (transversal shear modes).

Let $\delta y_{\beta}(\mathbf{r},t)=y_{\beta}(\mathbf{r},t)-y_{H\beta}(t)$ denote the deviation of $\{n, \mathbf{U}, T\}$ from their values in the HCS. Therefore, the hydrodynamic fields can be written as
\begin{subequations}
\begin{equation}
\label{4.3}
n(\mathbf{r},t)=n_H+\delta n (\mathbf{r},t), \quad \mathbf{U}(\mathbf{r},t)=\delta \mathbf{U}(\mathbf{r},t),
\eeq
\beq
T(\mathbf{r},t)=T_H(t)+\delta T (\mathbf{r},t),
\end{equation}
\end{subequations}
where the subscript $H$ means that the quantities are defined in the HCS. The hydrodynamic fields in the homogeneous state verify $\nabla n_H=\nabla T_H=0$ and the granular temperature $T_H$ is given by Eq.\ \eqref{4.2}. If the spatial perturbation is sufficiently small, then for some initial time interval these deviations will remain small and the hydrodynamic equations \eqref{3.25}--\eqref{3.27} can be linearized with respect to $\delta y_{\beta}(\mathbf{r},t)$. To recover the results found for dry granular fluids, we consider the same
time and space variables as those used in Refs.\ \cite{BDKS98,G05}, namely,
\begin{equation}
\label{4.4}
\tau=\frac{1}{2}\int_0^t\; \dd t'\; \nu_H(t'), \quad \boldsymbol{\ell}= \frac{1}{2}\frac{\nu_H(t)}{v_H(t)}\mathbf{r},
\end{equation}
where $\nu_H(t)$ is defined by Eq.\ \eqref{2.20} and $v_H(t)=\sqrt{T_H(t)/m}$. According to Eq.\ \eqref{4.4}, the unit length $\nu_H(t)/v_H(t)$ is proportional to the effective time-independent mean free path $1/n_H\sigma^{d-1}$. The dimensionless time scale $\tau$ is the time integral of the average collision frequency and thus is a measure of the average number of collisions per particle in the time interval between $0$ and $t$.

A set of Fourier transformed dimensionless variables are then
introduced by
\begin{equation}
\label{4.5}
\rho_{{\bf k}}(\tau)=\frac{\delta n_{{\bf k}}(\tau)}{n_{H}}, \quad
{\bf w}_{{\bf k}}(\tau)=\frac{\delta {\bf U}_{{\bf k}}(\tau)}{v_H(\tau)},\quad
\theta_{{\bf k}}(\tau)=\frac{\delta T_{{\bf k}}(\tau)}{T_{H}(\tau)},
\end{equation}
where $\delta y_{{\bf k}\alpha}\equiv \{\delta n_{{\bf k}},{\bf
w}_{{\bf k}}(\tau), \theta_{{\bf k}}(\tau)\}$ is defined as
\begin{equation}
\label{4.6}
\delta y_{{\bf k}\alpha}(\tau)=\int \dd {\boldsymbol {\ell}}\;
e^{-i{\bf k}\cdot {\boldsymbol {\ell}}}\delta y_{\alpha}
({\boldsymbol {\ell}},\tau).
\end{equation}
Note that in Eq.\ \eqref{4.6} the wave vector ${\bf k}$ is dimensionless.

As expected, the $d-1$ transverse velocity components ${\bf w}_{{\bf k}\perp}={\bf w}_{{\bf k}}-({\bf w}_{{\bf k}}\cdot
\widehat{{\bf k}})\widehat{{\bf k}}$ (orthogonal to the wave vector ${\bf k}$) decouple from the other  three modes and hence can be obtained more easily. Their evolution equation is
\begin{equation}
\label{4.7}
\frac{\partial {w}_{{\bf k}\perp}}{\partial \tau}=\left(2\gamma^*+\zeta_0^*-\frac{1}{2}\eta^*
k^2\right){w}_{{\bf k}\perp},
\end{equation}
where $\zeta_0^*\equiv \zeta_H^{(0)}/\nu_H$, $\gamma^*\equiv \gamma_H/\nu_H$ and $\eta^*\equiv \eta_H/\eta_{0H}$. All these quantities are evaluated in the reference base state (HCS). Since $\gamma^*$, $\zeta_0^*$, and $\eta^*$ are independent of time, then Eq.\ \eqref{4.7} is a simple (linear) differential equation whose solution is
\beq
\label{4.8}
{w}_{{\bf k}\perp}(\tau)={w}_{{\bf k}\perp}(0) \exp\left[\lambda_\perp(k) \tau\right],
\eeq
where
\beq
\label{4.9}
\lambda_\perp(k)=\zeta_0^*+2\gamma^*-\frac{1}{2}\eta^* k^2.
\eeq
Equation \eqref{4.9} identifies a critical wave number
\beq
\label{4.10}
k_s=\sqrt{\frac{2(2\gamma^*+\zeta_0^*)}{\eta^*}}
\eeq
such that the transversal shear modes always decay if $k>k_{s}$ while they grow exponentially when $k<k_{s}$. Therefore, when $k>k_s$ the transversal shear modes (velocity vortices) are linearly \emph{stable} while when $k<k_s$ the above modes are linearly \emph{unstable}.

The longitudinal three modes correspond to $\rho_{{\bf k}}$, $\theta_{{\bf k}}$, and the longitudinal velocity component of the velocity field, $w_{{\bf k}||}={\bf w}_{{\bf k}}\cdot \widehat{{\bf k}}$ (parallel to ${\bf k}$). These modes are coupled and obey the equation
\begin{equation}
\frac{\partial \delta y_{{\bf k}\alpha }(\tau )}{\partial \tau }=M_{\alpha \beta}
 \delta y_{{\bf k}\beta }(\tau ),
\label{4.11}
\end{equation}
where $\delta y_{{\bf k}\alpha }(\tau )$ denotes now the set $\left\{\rho_{{\bf k}}, \theta_{{\bf k}}, w_{{\bf k}||}\right\}$ and $\mathsf{M}$ is the square matrix
\begin{equation}
\mathsf {M}=\left(
\begin{array}{ccc}
0 & 0 & -i k \\
-2\left(\zeta_0^*g+2\gamma^*\right)-\displaystyle{\frac{d+2}{2(d-1)}}\mu^*k^2&\quad -\zeta_0^*-2\gamma^*-\displaystyle{
\frac{d+2}{2(d-1)}}\kappa^*k^2 & -ik(\frac{2}{d}p^*+\zeta_U)\\
-ikp^*C_\rho & -ikp^* &\zeta_0^{*}+2\gamma^*-\left(\displaystyle{\frac{d-1}{d}}\eta^*+\displaystyle{\frac{1}{2}}\eta_b^*\right)k^2
\end{array}
\right).   \label{4.12}
\end{equation}
As before, it is understood that $\eta^*$, $\eta_b^*$, $\kappa^*$, $\mu^*$,
$\zeta_0^*$, and $\zeta_U$ are evaluated in the HCS. Furthermore, the quantities $p^*(\alpha,\phi)$, $g(\phi)$, and
$C_\rho(\alpha,\phi)$ are given, respectively, by
\beq
\label{4.13}
p^*\equiv \frac{p_H}{n_H T_H}=1+2^{d-2}(1+\alpha)\chi \phi,
\eeq
\begin{equation}
\label{4.14}
g(\phi)=1+\phi\frac{\partial}{\partial \phi}\ln \chi(\phi),
\quad C_\rho(\alpha,\phi)=1+g(\phi)\frac{p^*(\alpha,\phi)-1}{p^*(\alpha,\phi)}.
\end{equation}
In the absence of the gas phase ($\gamma^*=0$), the matrix equation \eqref{4.12} is consistent with previous results derived for a dry three-dimensional granular fluid \cite{G05}. Moreover, for ordinary dilute gases ($\al=1$ and $\phi=0$), Eq.\ \eqref{4.12} also agrees with the results derived in Ref.\ \cite{PG14} for the stability analysis of a dilute gas subjected to a drag force.

The longitudinal three modes have the form $\text{exp}[\lambda_n (k)\tau]$ for $n=1,2,$ and $3$, where $\lambda_n (k)$ are the eigenvalues of the matrix $\mathsf{M}$, namely, they are the solutions of the cubic equation
\beq
\label{4.15}
\lambda^3+A \lambda^2+B \lambda+C=0,
\eeq
where
\beq
\label{4.16}
A(k)=\left[\frac{d+2}{2(d-1)}\kappa^*+\frac{d-1}{d}\eta^*+\frac{1}{2}\eta_b^*\right]k^2,
\eeq
\beqa
\label{4.17}
B(k)&=&\frac{d+2}{2d}\kappa^* \left[\eta^*+\frac{d}{2(d-1)}\eta_b^* \right]k^4+\left[p^* C_\rho
+p^*\left(\frac{2}{d}p^*+\zeta_U\right)+(2\gamma^*+\zeta_0^*)\right.\nonumber\\
& & \left.\times\left(
\frac{d-1}{d}\eta^*+\frac{1}{2}\eta_b^*-\frac{d+2}{2(d-1)}\kappa^*\right)\right]k^2-(2\gamma^*+\zeta_0^{*})^2,
\eeqa
\beq
\label{4.18}
C(k)=p^* \left[\frac{d+2}{2(d-1)}\left(\kappa^*C_\rho-\mu^*\right)k^2+\zeta_0^*
\left(C_\rho-2g\right)+2\gamma^*\left(C_\rho-2\right)\right]k^2.
\eeq

Before considering the general case $k \neq 0$, it is instructive to consider first the solutions to Eq.\ \eqref{4.15} in the extreme long wavelength limit, $k=0$ (Euler hydrodynamic order). In this case, the eigenvalues of the hydrodynamic modes are given by
\beq
\label{4.18.1}
\lambda_\perp=2\gamma^*+\zeta_0^*, \quad \lambda_{||}=\left\{0, -2\gamma^*-\zeta_0^*, 2\gamma^*+\zeta_0^*\right\}.
\eeq
Since two of the eigenvalues are positive (corresponding to growth of the initial perturbation in time), then some of the solutions are unstable. The zero eigenvalue represents a marginal stability solution, while the negative eigenvalue gives stable solutions. For general initial perturbations, all the modes are excited. These modes correspond to evolution of the granular suspension due to \emph{uniform} perturbations of the HCS, namely, a global change in the fields of the HCS. The unstable modes are seen to arise from the initial perturbations ${\bf w}_{{\bf k}\perp}(0)$ or ${\bf w}_{{\bf k}||}(0)$. These unstable modes may be considered as trivial since they appear due entirely to the normalization of the fluid velocity by the time dependent thermal velocity $v_H(t)$. However, this normalization is required to obtain time independent coefficients after scaling the set of fluid dynamic equations.

\begin{figure}
{\includegraphics[width=0.4\columnwidth]{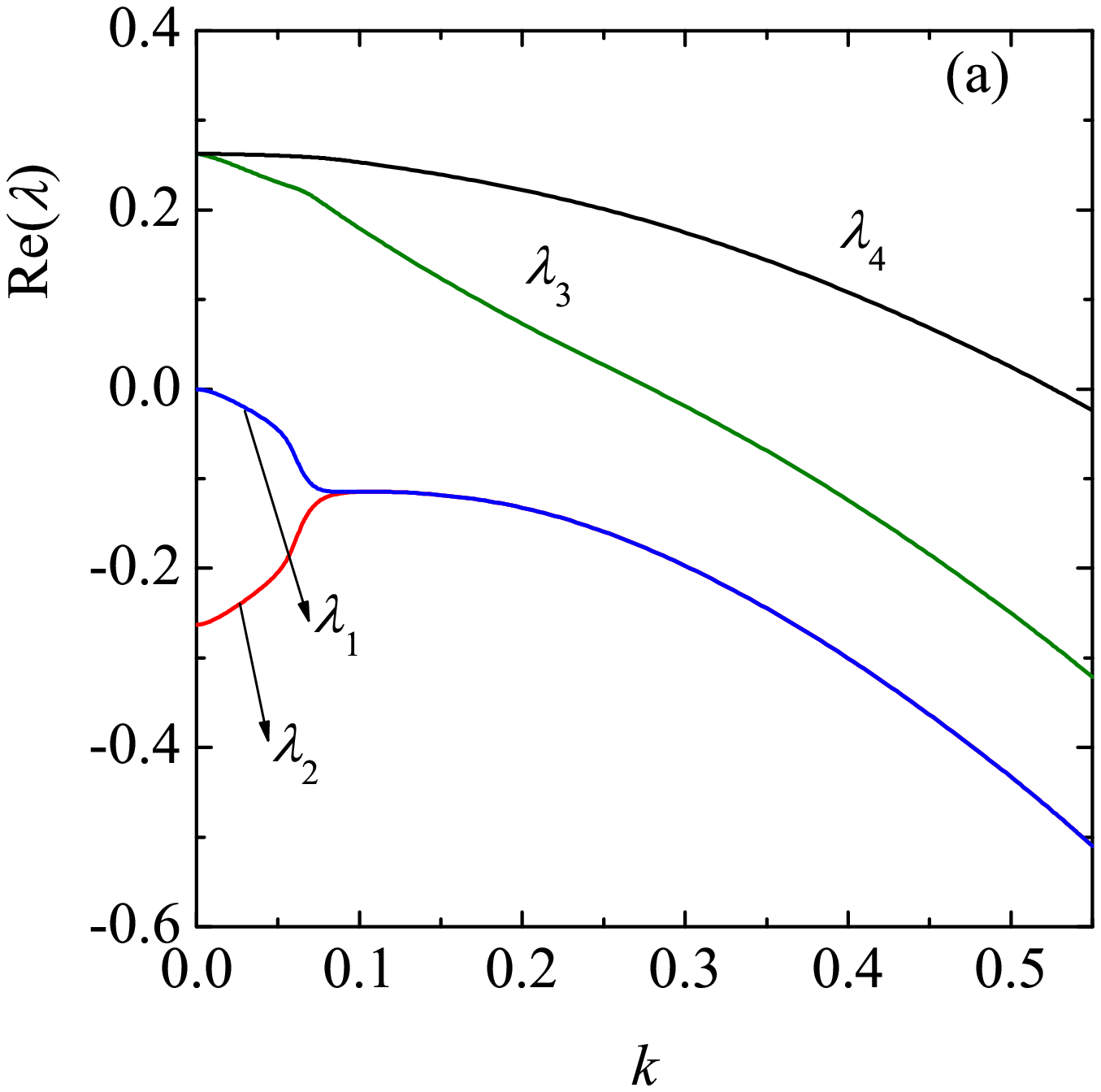}}
{\includegraphics[width=0.4\columnwidth]{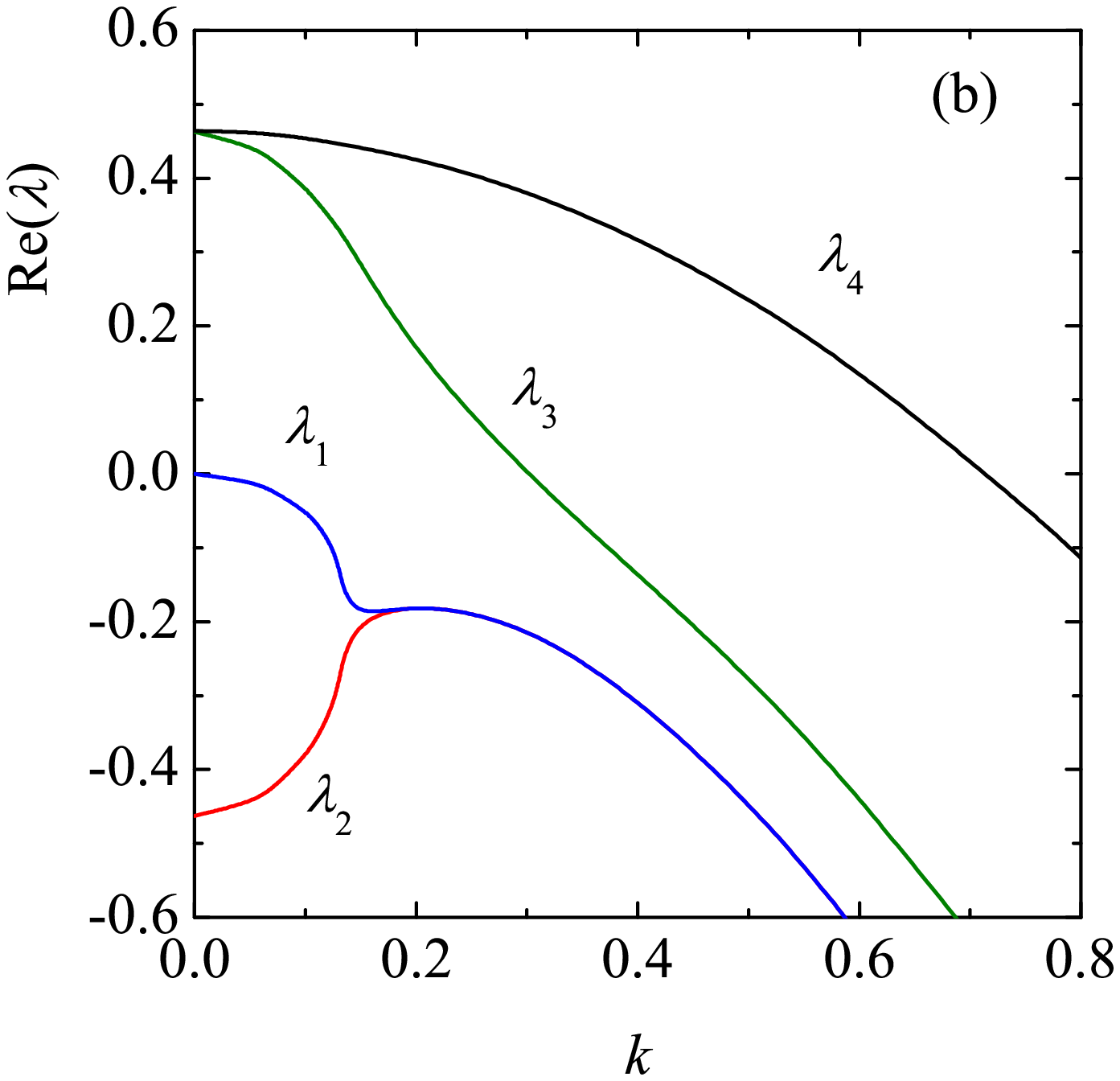}}
\caption{(Color online) (a) Dispersion relations for a dry three-dimensional granular fluid ($\gamma^*=0$) with $\al=0.8$ and $\phi=0.2$. From top to bottom, the curves correspond to the $d-1$ degenerate shear (transversal) modes and the remaining three longitudinal modes. Only the real parts of the eigenvalues is represented. (b) The same as in Fig.\ \ref{fig4}(a) but for a granular suspension with $\gamma^*=0.1$.
\label{fig4}}
\end{figure}
\begin{figure}
{\includegraphics[width=0.4\columnwidth]{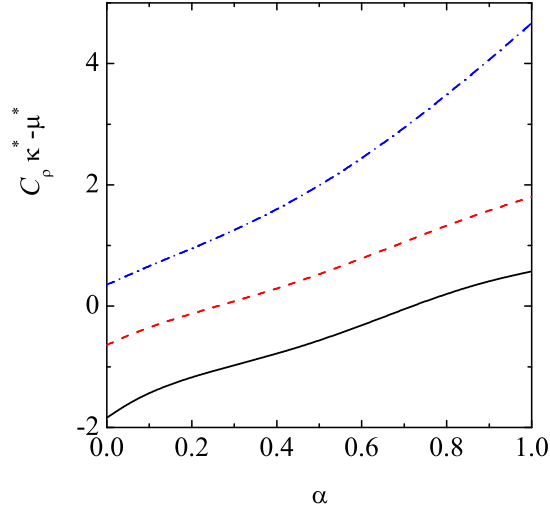}}
\caption{(Color online) Plot of the function $C_\rho \kappa^*-\mu^*$ versus the coefficient of restitution $\al$ for a three-dimensional system with $\gamma^*=0.2$ and three different values of the solid volume fraction $\phi$: $\phi=0$ (solid line), $\phi=0.1$ (dashed line), and $\phi=0.2$ (dash-dotted line).
\label{fig5}}
\end{figure}

For $k \neq 0$, the solution to Eq.\ \eqref{4.15} can be obtained for small wave numbers $k$ by a perturbation expansion as
\beq
\label{4.19}
\lambda_n(k)=\lambda_n^{(0)}+k \lambda_n^{(1)}+k^2 \lambda_n^{(2)}+\ldots,
\eeq
where the Euler eigenvalues $\lambda_n^{(0)}$ are given by the second identity of Eq.\ \eqref{4.18.1}. Substituting the expansion \eqref{4.19} into Eq.\ \eqref{4.15} yields
\beq
\label{4.21}
\lambda_n^{(1)}=0, \quad n=1,2,3,
\eeq
\beq
\label{4.22}
\lambda_1^{(2)}=p^*\frac{2(C_\rho-2)\gamma^*+\zeta_0^*(C_\rho-2g)}{(2\gamma^*+\zeta_0^*)^2},
\eeq
\beq
\label{4.23}
\lambda_2^{(2)}=-\frac{d+2}{2(d-1)}\kappa^*+\frac{p^*}{(2\gamma^*+\zeta_0^*)^2}\left[
\left(\frac{p^*}{d}+\frac{1}{2}\zeta_U\right)\left(2\gamma^*+\zeta_0^*\right)+2\gamma^*+g\zeta_0^*\right],
\eeq
\beqa
\label{4.24}
\lambda_3^{(2)}&=&-\eta^*-\frac{1}{2}\eta_b^*-\frac{p^*}{(2\gamma^*+\zeta_0^*)^2}\left\{\gamma^*\left(
2C_\rho-2+\zeta_U\right)+\zeta_0^*\left(C_\rho-g+\frac{1}{2}\zeta_U\right)\right.\nonumber\\
& & \left.-\frac{2\gamma^*+\zeta_0^*}{dp^*}\left[\left(2\gamma^*+\zeta_0^*\right)\eta^*-p^{*2}\right]
\right\}.
\eeqa
As before, when $\gamma^*=0$ and $d=3$, the expressions \eqref{4.21}--\eqref{4.24} agree with those previously derived for a dry granular fluid \cite{G05}. Since the Navier-Stokes hydrodynamic equations apply to second order in $k$, then the forms \eqref{4.19}--\eqref{4.24} are relevant to the same order.

The dispersion relations $\lambda_n(k)$ ($n=1,2,$ and $3$ correspond to the three longitudinal modes and $n=4$ is for the $d-1$ degenerate transversal shear modes) for a granular suspension with $\al=0.8$ and $\phi=0.2$, as obtained from Eq.\ \eqref{4.10} and the solutions of the cubic equation \eqref{4.15}, are plotted in Fig.\ \ref{fig4}. Two different values of the (reduced) friction coefficient have been considered: $\gamma^*=0$ (dry granular fluid) and $\gamma^*=0.1$. Only the real part (propagating modes) of the solutions to Eq.\ \eqref{4.15} have been represented. We observe that the dependence of $\text{Re}(\lambda)$ on the wave number $k$ is quite similar in both systems since while the solutions $\lambda_1$, $\lambda_2$, and $\lambda_3$ are real for small values of $k$, two of them ($\lambda_1$ and $\lambda_2$) become complex conjugate for wave numbers larger than a certain value. As expected, Fig.\ \ref{fig4} also shows that the heat mode ($\lambda_3$) is unstable for $k<k_h$, where $k_h$ can be determined from Eq.\ \eqref{4.15} when $s=0$. The result is
\beq
\label{4.25}
k_h=\sqrt{\frac{2(d-1)}{d+2}\frac{(2g-C_\rho)\zeta_0^*+2(2-C_\rho)\gamma^*}{C_\rho \kappa^*-\mu^*}}.
\eeq
Figure \ref{fig4} indicates that $k_s>k_h$ for the cases studied in this figure. Moreover, it appears that the effect of the interstitial gas on $k_s$ is to increase its value with respect to the one obtained in the dry granular case. On the other hand, while the numerator $(2g-C_\rho)\zeta_0^*+2(2-C_\rho)\gamma^*$ of Eq.\ \eqref{4.25} is always positive, the denominator $C_\rho \kappa^*-\mu^* >0$ can be negative. This is because the coefficient $\mu^*$ can be larger than the thermal conductivity $\kappa^*$ for extreme inelasticity, large values of $\gamma^*$ and/or very dilute systems (see Figs.\ \ref{kappa} and \ref{mu}). Consequently, the expression \eqref{4.25} is limited to values of $\al$, $\gamma^*$, and $\phi$ such that the combination $C_\rho \kappa^*-\mu^* >0$. To illustrate the dependence of $C_\rho \kappa^*-\mu^*$ on the coefficient of restitution $\al$, Fig.\ \ref{fig5} shows $C_\rho \kappa^*-\mu^*$ versus $\al$ for a relatively large value of the friction coefficient ($\gamma^*=0.2$) and three different values of density: a dilute gas ($\phi=0)$ and two moderately dense systems ($\phi=0.1$ and 0.2). While the combination $C_\rho \kappa^*-\mu^*$ is always positive for any value of $\al$ for $\phi=0.2$  (and hence, $k_s$ is well defined in the complete range of values of $\al$), the above combination turns out to be negative for very dilute systems when $\al$ is smaller than a certain value $\al'(\phi)$. In particular, $\al'\simeq 0.26$ for $\phi=0.1$ and $\al'\simeq 0.72$ for $\phi=0$. This means that while Eq.\ \eqref{4.25} is restricted to a small region of the parameter space for very dilute systems, the above expression for $k_h$ can be applied for both arbitrary values of $\al$ and relatively large values of $\gamma^*$ in the case of moderately dense systems. It is also important to recall that Eq.\ \eqref{4.25} is always well defined in the dry granular case ($\gamma^*=0$) and hence, its limitations are only restricted to very \emph{dilute} granular suspensions. In any case, one has to note that usually clustering instability appears at a time where the spatial gradients are not small and so, a linear stability analysis (which neglects nonlinear terms in perturbations) is not expected to match well the results obtained in computer simulations for density clusters. In particular, one of these terms (the viscous heating term $P_{ij}\partial_j U_i$) has been shown to play an important role in the detection of clustering instability by means of hydrodynamic theories \cite{BRC99} and particle simulations \cite{SMM00,MDHEH13} in dry granular fluids. Similar trends are expected in granular gas-solid flows.

\begin{figure}
{\includegraphics[width=0.4\columnwidth]{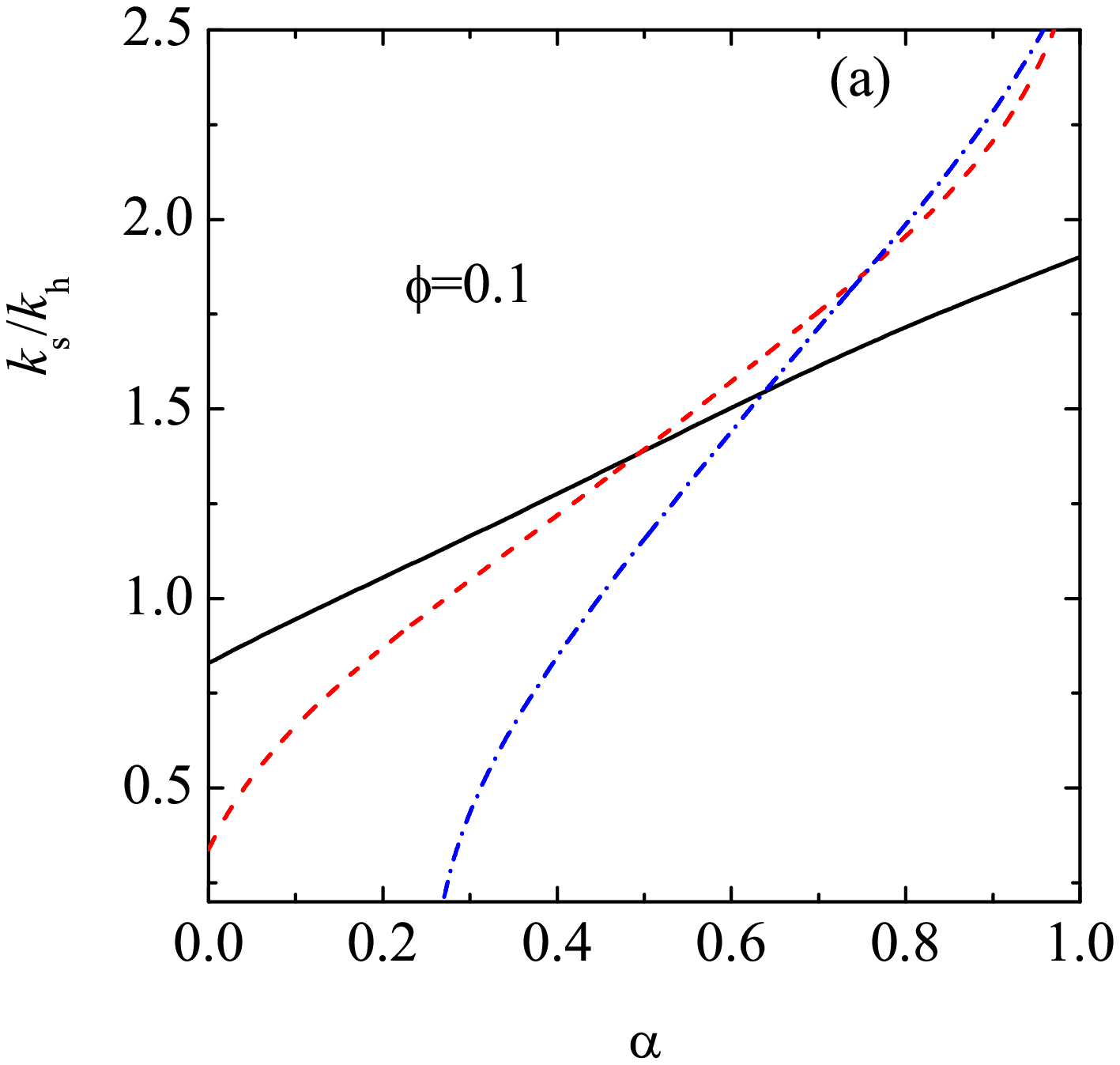}}
{\includegraphics[width=0.4\columnwidth]{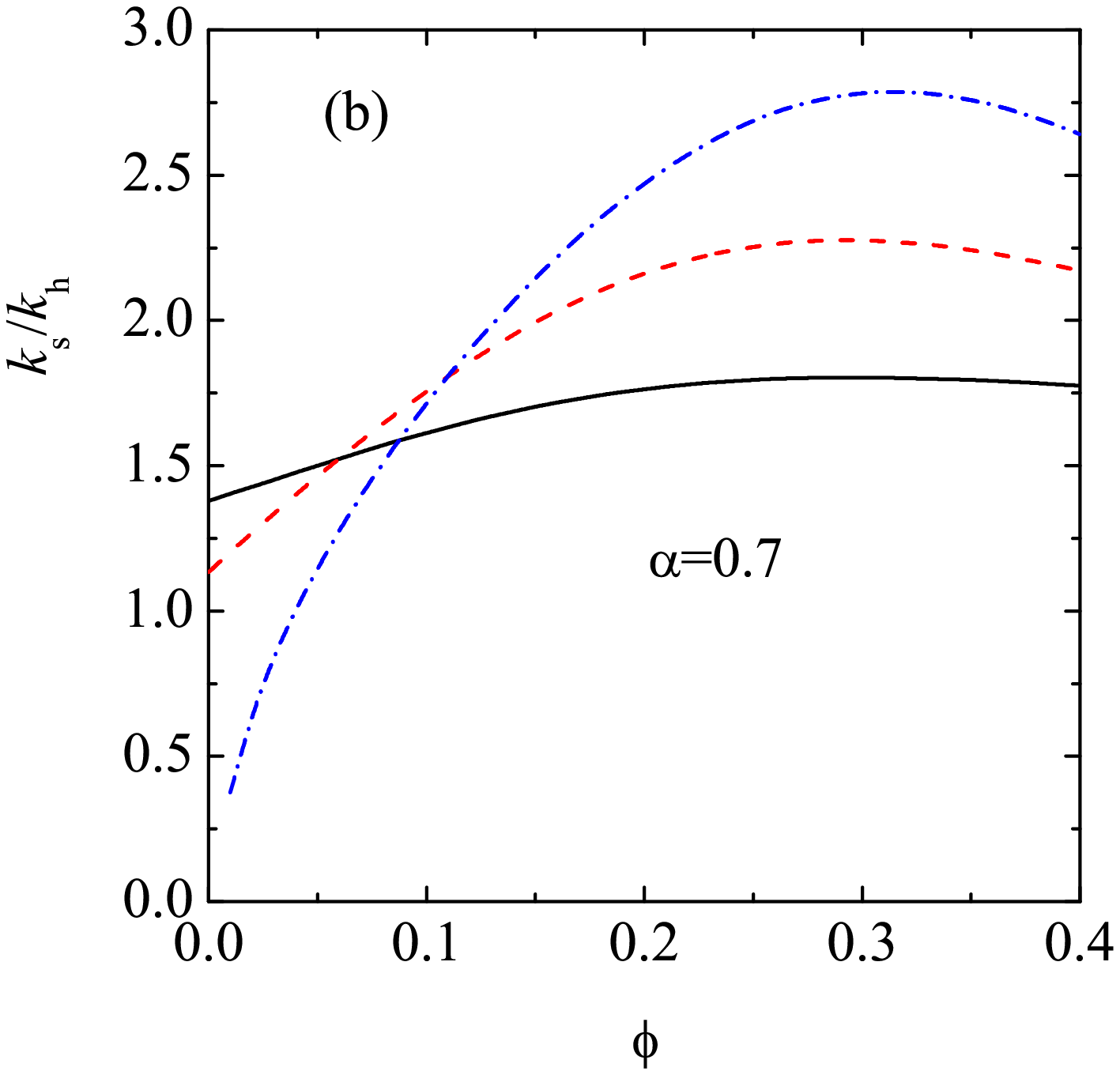}}
\caption{(Color online) (a) Ratio $k_s/k_h$ versus the coefficient of restitution $\al$ for a three-dimensional system with $\phi=0.1$ and three values of the (reduced) friction coefficient $\gamma^*$: $\gamma^*=0$ (solid line), $\gamma^*=0.1$ (dashed line), and $\gamma^*=0.2$ (dash-dotted line). (b) Ratio $k_s/k_h$ versus the solid volume fraction $\phi$ for a three-dimensional system with $\al=0.7$ and three values of the (reduced) friction coefficient $\gamma^*$: $\gamma^*=0$ (solid line), $\gamma^*=0.1$ (dashed line), and $\gamma^*=0.2$ (dash-dotted line).
\label{fig6}}
\end{figure}
\begin{figure}
{\includegraphics[width=0.4\columnwidth]{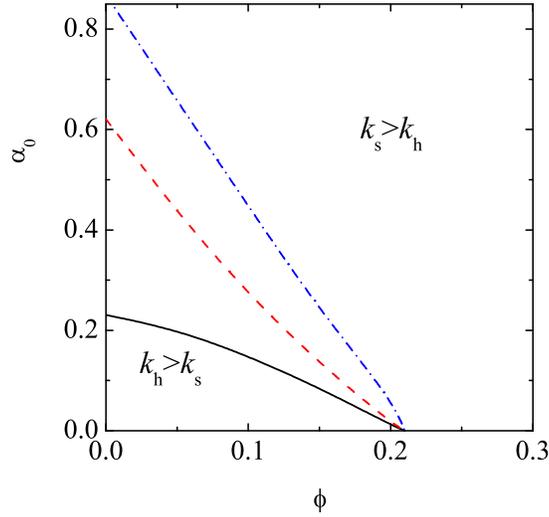}}
\caption{(Color online) Dependence of $\al_0$ on the solid volume fraction $\phi$ for a three-dimensional system with three values of the (reduced) friction coefficient $\gamma^*$: $\gamma^*=0$ (solid line), $\gamma^*=0.1$ (dashed line), and $\gamma^*=0.2$ (dash-dotted line). Points above (below) the curve correspond to systems where the instability is driven by the shear (heat) mode.
\label{fig7}}
\end{figure}

The ratio $k_s/k_h$ measures the separation between the critical shear and heat modes. As said before, while the critical wave number $k_s$ is associated with velocity vortices, $k_h$ is related with density clusters. It is quite apparent that the ratio $k_s/k_h$ presents a complex dependence on the parameter space of the system. To illustrate the $\al$-dependence of $k_s/k_h$, Fig.\ \ref{fig6}(a) shows $k_s/k_h$ versus $\al$ for a moderately three-dimensional dense fluid ($\phi=0.1$). Three values of $\gamma^*$ have been considered. It appears that in general both critical modes $k_s$ and $k_h$ are well separated, especially for small inelasticity. We also observe that for not quite strong dissipation, the critical shear mode $k_s$ is larger than the critical heat mode $k_h$ and hence, the instability is driven by the former one. However, for a given value of $\gamma^*$, there exists a value of the coefficient of restitution $\al_0(\phi)$ for which $k_h>k_s$ for values of $\al<\al_0$. A remarkable point is that the value of $\al_0$ increases with $\gamma^*$ so that, the region where clusters are developed before than vortices becomes larger as the impact of the gas phase on the dynamics of grains is more significant. As a complement, the ratio $k_h/k_s$ is plotted versus $\phi$ for $\al=0.7$ in Fig.\ \ref{fig6}(b) for three values of $\gamma^*$. We observe that for dense systems (say for instance, $\phi \gtrsim 0.2$) the instability of the system is clearly driven by the transversal shear mode, regardless of the influence of the interstitial gas. On the other hand, for very dilute systems and for $\gamma^* \neq 0$, there are small regions in the parameter space of the system where the onset of instability is associated with the heat mode. The size of the latter region increases with $\gamma^*$.

Since the value of the coefficient of restitution plays a relevant role in the stability of the system, Fig.\ \ref{fig7} shows $\al_0$ versus the solid volume fraction $\phi$ for three values of $\gamma^*$. Figure \ref{fig7} highlights that in general the value of $\al_0$ decreases with increasing density. In particular, $k_s>k_h$ for $\phi \gtrsim 0.2$. On the other hand, as expected, the value of $\al_0$ increases as the density of the system decreases and/or the friction coefficient $\gamma^*$ increases. Thus, for $\gamma^*=0.2$, $\al_0 \simeq 0.86$ for a dilute gas ($\phi=0$). However, given that the values of $\al_0$ are quite small for moderate densities, one can conclude that in practice the onset of instability is usually dominated by the transversal shear mode (vortex instability). 

According to these results, for not quite strong values of dissipation, three different regions can be identified. For $k>k_s$, all the modes are negative and the system is linearly stable with respect to long enough wavelength excitations. For $k_h<k<k_s$, the shear mode is unstable while the heat mode is linearly stable. In this region, density inhomogeneities can only be created due to the nonlinear coupling with the unstable shear mode \cite{BRC99}. Finally, if $k<k_h$ first vortices and then clusters appear and the final state is strongly inhomogeneous.

\begin{figure}
{\includegraphics[width=0.4\columnwidth]{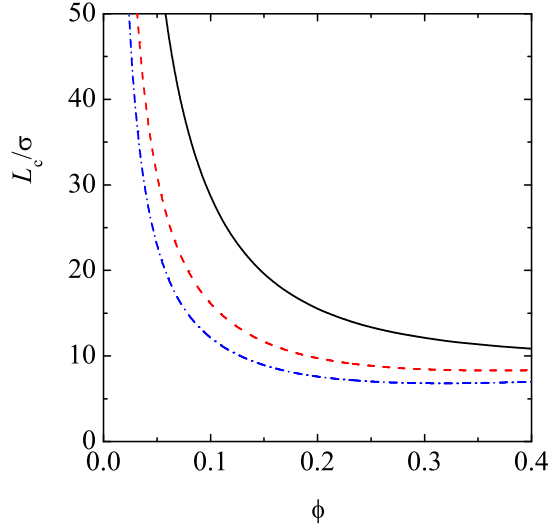}}
\caption{(Color online) The critical size in units of the diameter of spheres as a function of the solid volume fraction
$\phi$ for $\al=0.9$ and three values of the (scaled) friction coefficient $\gamma^*$:
$\gamma^*=0$ (solid line), $\gamma^*=0.1$ (dashed line), and $\gamma^*=0.2$ (dash-dotted line).
\label{fig8}}
\end{figure}

In a system of finite size with periodic boundary conditions, the smallest allowed wave number is $2\pi/L$, where $L$ is the largest system length. Hence, for given values of inelasticity, density, and viscous friction, a critical length $L_c$ can be identified and so, the system becomes unstable when its length $L>L_c$. In dimensionless units, the value of $L_c$ is
\beq
\label{4.26}
\frac{2\pi}{L_c^*}=\text{max}\left\{k_s,k_h\right\}, \quad L_c^*=\frac{\nu_H}{2v_H}L_c.
\eeq
The critical length $L_c$ (in units of the diameter $\sigma$) is plotted in Fig.\ \ref{fig8} as a function of the solid volume fraction $\phi$ for a three-dimensional system with $\al=0.9$ and three different values of the coefficient of friction $\gamma^*$. In all of these systems, $k_s>k_h$ and so
\beq
\label{4.27}
\frac{L_c}{\sigma}=\frac{(d+2)\pi^{3/2}}{2^d d \phi}\sqrt{\frac{\eta^*}{2\gamma^*+\zeta_0^*}}.
\eeq
We observe first that the inclusion of the interstitial gas gives rise to a systematic reduction in the critical length scale. This behavior agrees qualitatively well with the trends observed in the direct numerical simulations (DNS) reported in Fig.\ 10 of Ref.\ \cite{GFHY16}). In addition, in an attempt to make a quantitative comparison with the simulation data provided in Table I of Ref.\ \cite{GFHY16}, one should note first that the (scaled) friction coefficient $\gamma^*$ introduced in this work depends on time as
\beq
\label{4.28}
\gamma^*(t^*)=\frac{\gamma_0^*}{\sqrt{T(t^*)/T_0}},
\eeq
where $t^*=\nu(T_0) t$ is a dimensionless time, $T_0$ is the initial temperature and $\nu$ is defined in Eq.\ \eqref{2.20}. In terms of $t^*$, the evolution equation for the temperature is given by Eq.\ (33) of Ref.\ \cite{GFHY16}. Moreover, the (dimensionless) coefficient $\gamma_0^*$ can be written in terms of typical dimensionless numbers of suspensions, such as the ratio of the material densities of the solid and gas phases $\rho_s/\rho_g$ and the Reynolds number $\text{Re}_{T_0}$ based on the initial temperature $T_0$, namely,
\beq
\label{4.29}
\text{Re}_{T_0}=\frac{\sigma \rho_g}{\mu_g}\sqrt{\frac{T_0}{m}},
\eeq
where we recall that $\mu_g$ is the viscosity of gas phase. For hard spheres ($d=3$), $\gamma_0^*$ can be written as \cite{GFHY16}
\beq
\label{4.30}
\gamma_0^*=\frac{15}{16}\frac{\sqrt{\pi}}{\phi}\frac{\rho_g}{\rho_s}\frac{R_{\text{diss}}(\phi)}{\text{Re}_{T_0}},
\eeq
where $\rho_s=6m/\pi \sigma^3$ for spheres and $R_\text{diss}$ is \cite{SMTK96}
\beq
\label{4.31}
R_{\text{diss}}(\phi)=1+3\sqrt\frac{\phi}{2}+\frac{135}{64}\phi \ln
\phi+11.26 \phi \left(1-5.1 \phi+16.57 \phi^2-21.77 \phi^3\right)-\phi \chi(\phi)\ln 0.01.
\eeq
The simulations carried out in Ref.\ \cite{GFHY16} correspond to granular suspensions with $\text{Re}_{T_0}\approx 5$, $\rho_s/\rho_g=1000$ and two different values of the coefficient of restitution ($\al=0.9$ and 0.8). Considering a sufficiently long time ($t^*=200$) for which the hydrodynamic regime has been achieved, the values of $\gamma^*$ obtained from Eqs.\ \eqref{4.28} and \eqref{4.30} for $\al=0.9$ are $0.18$ and $0.24$ for the densities $\phi=0.2$ and 0.3, respectively. In the case $\al=0.8$, the values of $\gamma^*$ are 0.32 and 0.43 for the densities $\phi=0.2$ and 0.3, respectively. The corresponding theoretical predictions for $L_c/\sigma$ obtained from Eq.\ \eqref{4.27} by employing the above values of $\gamma^*$ for $\al=0.9$ are 7.92 and 6.38 for $\phi=0.2$ and 0.3, respectively, while the simulation data for the above densities are 6.54 and 5.55, respectively. In the case $\al=0.8$, the theory predicts $L_c/\sigma=5.60$ and 4.49 for the above two densities while the values of the critical system size obtained from the simulations are 4.48 and 4.03, respectively. Thus, as the theoretical results derived in Ref.\ \cite{GFHY16} when the time dependence of $\gamma^*$ is accounted for, the present theory overestimates the results obtained from simulations. On the other hand, the largest discrepancy between theory and simulations is smaller than 25\%.

\section{Drag force as a thermostatic force. Steady states}
\label{sec5}

\begin{figure}
{\includegraphics[width=0.4\columnwidth]{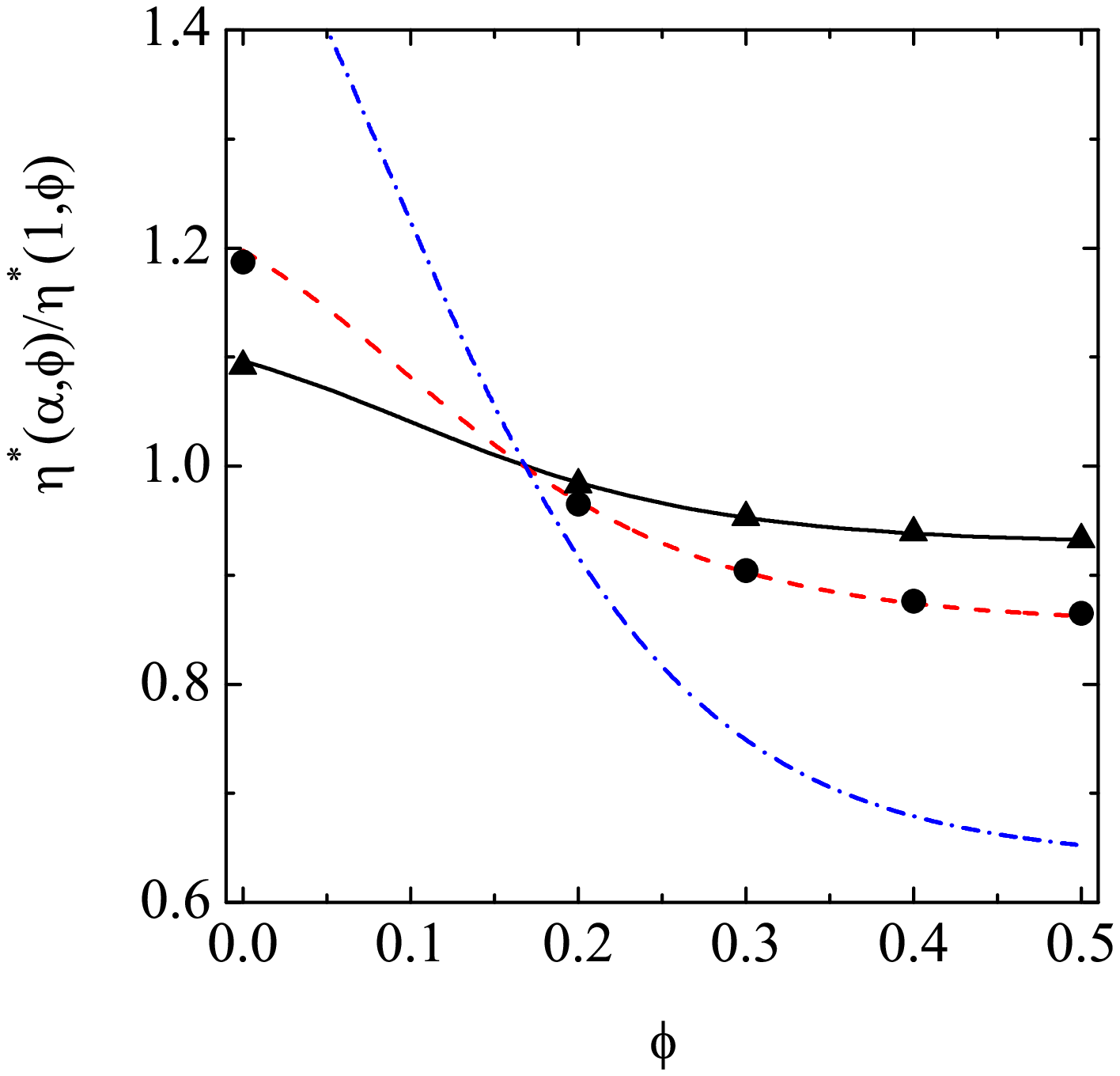}}
{\includegraphics[width=0.4\columnwidth]{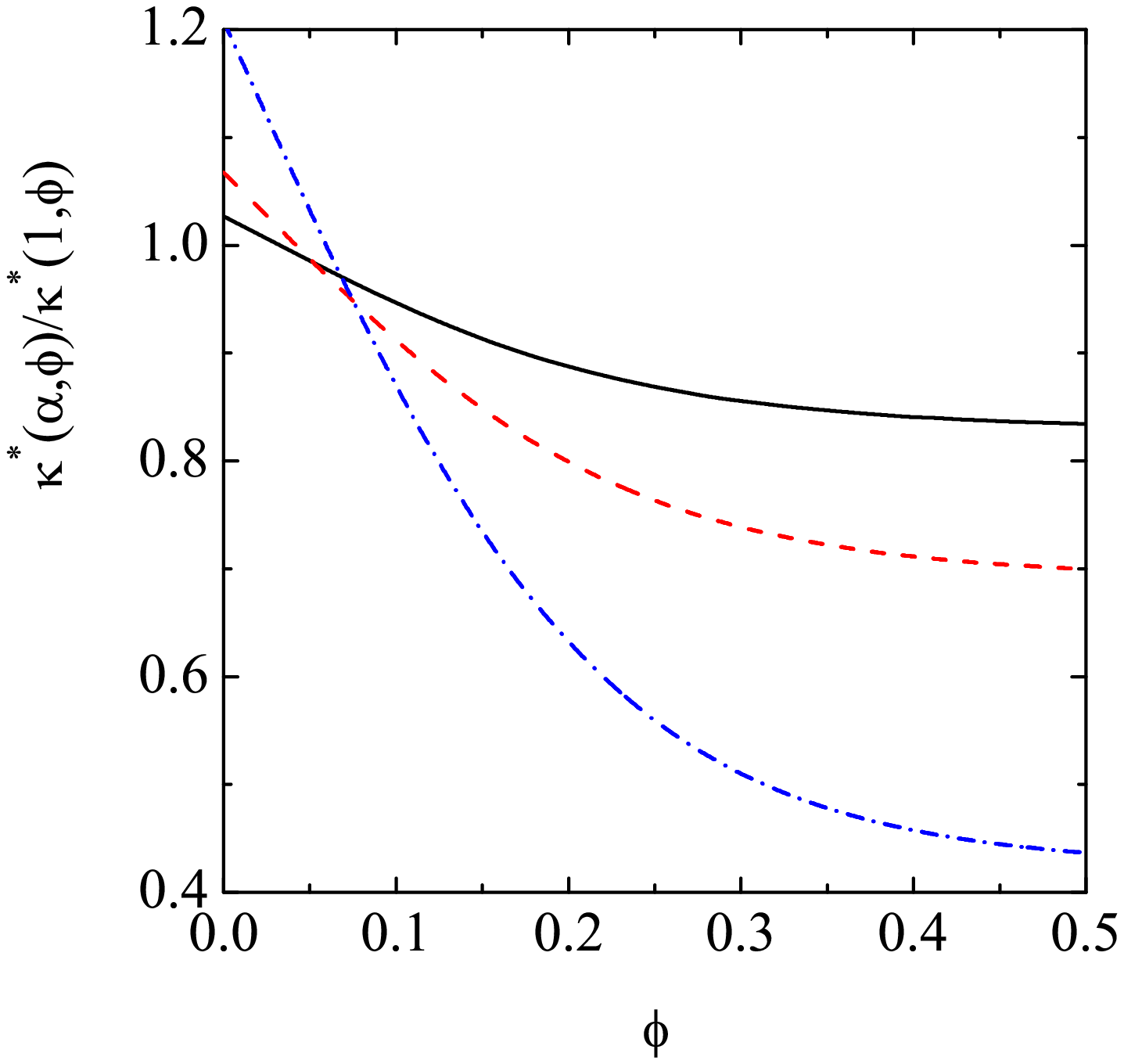}}
\resizebox{6.5cm}{!}{\includegraphics{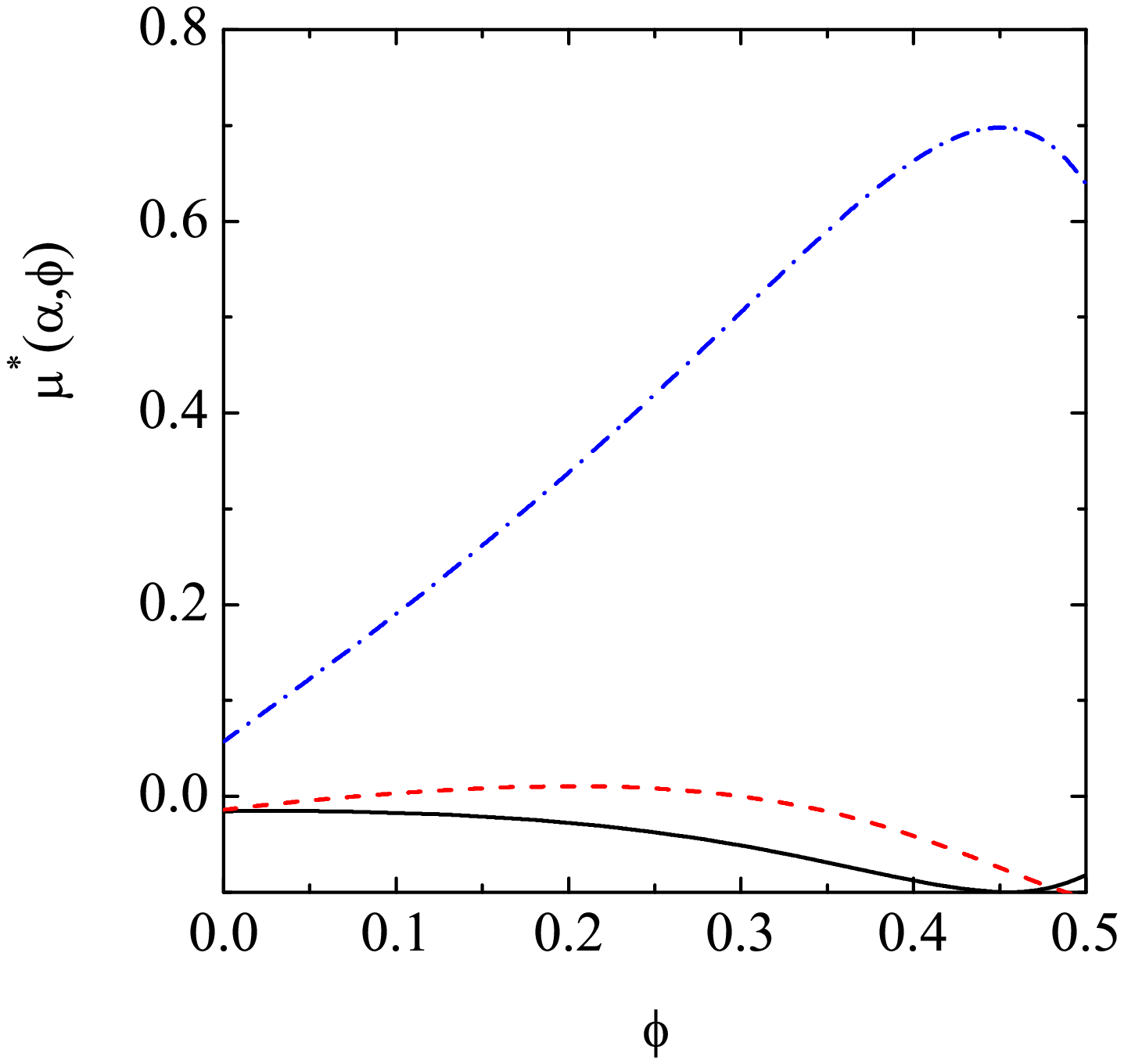}}
\caption{(Color online) Plot of the the steady (dimensionless) transport coefficients $\eta^*(\al,\phi)/\eta^*(1,\phi)$, $\kappa^*(\al,\phi)/\kappa^*(1,\phi)$, and $\mu^*(\al,\phi)$ as functions of the solid volume fraction $\phi$ for a three-dimensional system with three different values of the coefficient of restitution $\al$:
$\al=0.9$ (solid line), $\al=0.8$ (dashed line), and $\al=0.5$ (dash-dotted line). Symbols (triangles for $\al=0.9$ and circles for $\al=0.8$) refer to Monte Carlo simulations results obtained in Ref.\ \cite{GM03}.
\label{fig9}}
\end{figure}

In computer simulations \cite{EH86,EM90}, the drag force \eqref{1.1} has been widely employed to control the temperature of the system. Thus, apart from modeling the influence of the gas phase on the solid particles, the external force \eqref{1.1} can be considered as a thermostatic force to achieve steady states. In particular, for sheared molecular fluids ($\al=1$), the friction coefficient $\gamma$ is a positive shear-rate dependent function chosen to compensate for the viscous heating produced by shear work \cite{EM90,GS03}. On the other hand, in the case of granular fluids in homogeneous states, when $\gamma<0$ the system is heated by an ``antidrag'' force chosen to exactly compensate for collisional cooling and reach a non-equilibrium steady state \cite{MS00}. According to Eq.\ \eqref{4.1}, the condition $\partial_t T=0$ yields $\gamma=-\frac{1}{2}\zeta^{(0)}$ and hence,
\beq
\label{5.1}
\gamma(\phi,\al)=-\frac{d+2}{4d}\chi(\phi)(1-\al^2)\left(1+\frac{3}{16}a_2\right).
\eeq
Upon deriving the expression \eqref{5.1} use has been made of Eq.\ \eqref{a16}. Therefore, the friction coefficient $\gamma$ is taken as a negative coefficient coupled to the coefficient of restitution $\al$.

In the steady state, the expressions of the (scaled) transport coefficients $\eta^*$, $\eta_b^*$, $\kappa^*$, and $\mu^*$ can be easily derived from their general forms when $\gamma$ is replaced by its $\al$-dependent form \eqref{5.1}. Their expressions are given by Eqs.\ \eqref{3.3}--\eqref{3.6} where their kinetic contributions are
\beq
\label{5.2}
\eta_k^*=\frac{1-\displaystyle{\frac{2^{d-2}}{d+2}}(1+\al)(1-3\al)\phi \chi}{\nu_\eta^*-\zeta_0^*},
\eeq
\beq
\label{5.3}
\kappa_k^*=\frac{d-1}{d}\left(\nu_\kappa^*-\frac{3}{2}\zeta_0^*\right)^{-1}\left\{1+2a_2+3\frac{2^{d-3}}{d+2}\phi \chi
(1+\alpha)^2\left[2\alpha-1+a_2(1+\alpha)\right]\right\},
\eeq
\beqa
\label{5.4}
\mu_{k}^*&=&\left(\nu_\kappa^*-\frac{3}{2}\zeta_0^*\right)^{-1}\left\{\frac{d-1}{d}a_2
+\phi\frac{\partial \ln \chi}{\partial \phi}\zeta_0^{*} \kappa_k^*
+ 3\frac{2^{d-2}(d-1)}{d(d+2)}\phi \chi
(1+\alpha)\left(1+\frac{1}{2}\phi\frac{\partial \ln \chi}{\partial \phi}\right)\right.\nonumber\\
& & \left.\times
\left[\alpha(\alpha-1)+\frac{a_2}{6}(10+2d-3\alpha+3\alpha^2)\right]\right\},
\eeqa
When $\phi=0$, Eqs.\ \eqref{5.2}--\eqref{5.4} are consistent with those derived for granular dilute gases driven by a Gaussian thermostat \cite{GM02}.

The dependence of the (steady) transport coefficients $\eta^*$, $\kappa^*$, and $\mu^*$ on the solid volume fraction is plotted in Fig.\ \ref{fig9} for three different values of the coefficient of restitution $\al$. The theoretical expressions for the shear viscosity are also compared with Monte Carlo simulations of the Enskog equation carried out in Ref.\ \cite{GM03}. We observe first that the theoretical predictions for $\eta^*$ show an excellent agreement with simulations, even for strong dissipation and high densities. Both theory and simulation show
that, at a given value of $\al$, $\eta^*(\al,\phi)>\eta^*(1,\phi)$ if the solid volume fraction $\phi$ is smaller than a certain value $\phi_0$, while $\eta^*(\al,\phi)<\eta^*(1,\phi)$ if $\phi>\phi_0$. Thus, while for $\phi<\phi_0$ the shear viscosity $\eta^*(\al,\phi)$ increases with $\al$, the opposite happens for sufficiently dense fluids ($\phi>\phi_0$). A similar behavior is found for the (scaled) thermal conductivity $\kappa^*(\al,\phi)$, although the valor of $\phi_0$ is smaller than the one obtained for $\eta^*$. As a consequence, only for relatively dilute granular gases ($\phi \lesssim 0.05$) the thermal conductivity increases with inelasticity. Regarding the dependence of $\mu^*$ on $\phi$, it is quite apparent that $\mu^* \simeq 0$ for not quite strong dissipation in the limit of very dilute gases ($\phi \simeq 0$). This is basically due the fact that $\mu^* \propto a_2(\al)$ which is in general very small. This means that, for practical purposes, one can neglect the contribution to the heat flux coming from the term proportional to the density gradient and so, $\mathbf{q}^{(1)}\to -\kappa \nabla T$. On the other hand, the value of $\mu^*$ considerably increases at extreme inelasticity (say for instance, $\al \sim 0.5$) especially for dense systems. In fact, its magnitude can be even larger than that of the thermal conductivity $\kappa$ for this regime of densities.

\section{Concluding remarks}
\label{sec6}

This paper addresses the effect of the interstitial gas on the instabilities of the HCS of an assembly of smooth, inelastic, monodisperse solid particles. A viscous drag force term for the gas phase is incorporated into the Enskog kinetic equation to account for the influence of the former on the transport properties of grains. This model (thermal drag model) can be seen as a simplified version of a more complete model \cite{GTSH12} where the mean relative velocity between the solid and gas phases can be neglected. A previous analysis on the stability of the HCS in granular gas-solid flows has been recently \cite{GFHY16} carried out by using the thermal drag model. In this work \cite{GFHY16}, the complete granular temperature dependence of the scaled friction coefficient $\gamma^*$ was accounted for in the determination of the Navier-Stokes transport coefficients. As a consequence and unlike the dry granular stability analysis \cite{BDKS98,G05}, all the (scaled) hydrodynamic modes involved in the analysis performed in Ref.\ \cite{GFHY16} become functions of time through the scaled friction coefficient $\gamma^*$ (which is proportional to $T(t)^{-1/2}$). Thus, this stability analysis required a numerical solution and in particular, only the time dependence of the transversal shear modes (velocity vortices) was studied by using a simple method proposed to determine the neutral stability.

A possible way of offering an analytical treatment of the linear stability analysis of the HCS in granular suspensions is to assume that the (scaled) friction coefficient $\gamma^*$ is constant. This is equivalent to chose the coefficient $\gamma$ proportional to the effective collision frequency $\nu$ for hard spheres. With this choice, all the scaled quantities involved in the stability analysis are independent of time after changing space and time [see Eq.\ \eqref{4.4}] and scaling the hydrodynamic fields [see Eq.\ \eqref{4.5}]. This allows us to determine analytically all the hydrodynamic modes and the critical length $L_c$ beyond which the system becomes unstable. In particular, as for dry granular gases, the results show that for dense gases the most unstable modes are the transversal shear modes. This means that velocity vortices precede to density clusters. However, this tendency changes as the density of the granular suspension decreases and/or the value of the (scaled) coefficient $\gamma^*$ increases [see Figs.\ \ref{fig6} and \ref{fig7}]. Regarding the impact of the gas phase on the critical length, for given values of the coefficient of restitution and density, our results indicate that the effect of the interstitial gas is to decrease the value of $L_c$ with respect to its dry granular value. This conclusion agrees qualitatively well with the numerical results obtained in Ref.\ \cite{GFHY16}. Moreover, at a more quantitative level, a comparison between our theoretical predictions for the vortex instability with those obtained from computer simulations \cite{GFHY16} (in the time window where our results could be applicable) indicates that our theory overestimates the simulation data. These discrepancies are smaller than 25 \% and 15\% for the densities $\phi=0.2$ and 0.3, respectively. As noted in Ref.\ \cite{GFHY16}, given that the quantitative agreement is not as favorable as previous studies have found for similar (dry) granular systems \cite{MDCPH11,MGHEH12,MGH14}, one should try to refine the theory to achieve more accurate results. Work along this line will be worked out in the next future. Finally, an extension of the results derived in this paper to the important subject of granular mixtures could be also an interesting project for the future.

\acknowledgments

The research of V. G. has been supported by the Spanish Government through grant No. FIS2013-42840-P, partially financed by FEDER funds and by the Junta de Extremadura (Spain) through Grant No. GR15104.

\appendix
\section{Kinetic contributions to the fluxes}
\label{appA}
The form of the first-order distribution $f^{(1)}$ is  given by
\beq
\label{a1}
f^{(1)}=\boldsymbol{\mathcal{A}}\left(\mathbf{V}\right)\cdot  \nabla \ln
T+\boldsymbol{\mathcal{B}}\left(
\mathbf{V}\right) \cdot \nabla \ln n
+\mathcal{C}_{ij}\left( \mathbf{V} \right)\frac{1}{2}\left( \partial _{i}U_{j}+\partial _{j
}U_{i}-\frac{2}{d}\delta _{ij}\nabla \cdot
\mathbf{U} \right)+\mathcal{D}\left( \mathbf{V} \right) \nabla \cdot
\mathbf{U},
\eeq
where the quantities $\boldsymbol{\mathcal{A}}\left(\mathbf{V}\right)$, $\boldsymbol{\mathcal{B}}\left(
\mathbf{V}\right)$, $\mathcal{C}_{ij}\left( \mathbf{V} \right)$ and $\mathcal{D}\left( \mathbf{V} \right)$
are the solutions of the following linear integral equations:
\beq
\label{a2}
\frac{1}{2}\left(2\gamma +\zeta^{(0)}\right)\frac{\partial}{\partial \mathbf{V}}\cdot \left(
\mathbf{V}\boldsymbol{\mathcal{A}}\right)-\left(\gamma+\frac{1}{2}\zeta^{(0)}\right)
\boldsymbol{\mathcal{A}}-\gamma\frac{\partial}{\partial {\bf V}}\cdot {\bf V}\boldsymbol{\mathcal{A}}
+{\cal L}\boldsymbol{\mathcal{A}}={\bf A},
\eeq
\begin{equation}
\label{a3}
\frac{1}{2}\left(2\gamma +\zeta^{(0)}\right)\frac{\partial}{\partial \mathbf{V}}\cdot \left(
\mathbf{V}\boldsymbol{\mathcal{B}}\right)-\gamma\frac{\partial}{\partial {\bf V}}\cdot {\bf V}\boldsymbol{\mathcal{B}}
+{\cal L}\boldsymbol{\mathcal{B}}
={\bf B}+\left[2\gamma +\zeta ^{(0)}\left(1+\phi
\frac{\partial \ln \chi}{\partial \phi} \right)\right]\boldsymbol{\mathcal{A}},
\end{equation}
\begin{equation}
\label{a4}
\frac{1}{2}\left(2\gamma +\zeta^{(0)}\right)\left[\mathcal{C}_{ij}+\frac{\partial}{\partial \mathbf{V}}\cdot
\left(\mathbf{V}\mathcal{C}_{ij}\right)\right]
-\gamma\frac{\partial}{\partial {\bf V}}\cdot {\bf V}\mathcal{C}_{ij}+{\cal L}\mathcal{C}_{ij}=C_{ij},
\end{equation}
\begin{equation}
\label{a5}
\frac{1}{2}\left(2\gamma +\zeta^{(0)}\right)\left[\mathcal{D}+\frac{\partial}{\partial \mathbf{V}}\cdot
\left(\mathbf{V}\mathcal{D}\right)\right]
-\gamma\frac{\partial}{\partial {\bf V}}\cdot {\bf V}\mathcal{D}+{\cal L}\mathcal{D}=D,
\end{equation}
where
\begin{equation}
\label{a6}
\mathcal{L}X=-\left(J_\text{E}^{(0)}[f^{(0)},X]+J_\text{E}^{(0)}[X,f^{(0)}]\right),
\end{equation}
and
\beq
\label{a7}
J_\text{E}^{(0)}[f^{(0)},f^{(0)}]=\chi \sigma^{d-1}\int\; \dd{\bf v}_{2}\int \dd\widehat{\boldsymbol{\sigma}}\,
\Theta (\widehat{{\boldsymbol {\sigma}}}\cdot {\bf g}_{12})(\widehat{\boldsymbol {\sigma }}\cdot {\bf g}_{12})\left[
\al^{-2}f^{(0)}(v_1') f^{(0)}(v_2')-f^{(0)}(v_1) f^{(0)}(v_2)\right]
\eeq
is the Boltzmann collision operator for inelastic collisions multiplied by the (constant) pair correlation function $\chi$. The inhomogeneous terms $\mathbf{A}$, $\mathbf{B}$, $C_{ij}$, and $D$ are given by Eqs.\ (A5)--(A8), respectively, of Ref.\ \cite{GTSH12}.

As noted in Ref.\ \cite{GFHY16}, the forms of the collisional contributions to the Navier-Stokes transport coefficients do not explicitly depend on the friction coefficient $\gamma$ and hence, their expressions are exactly the same as those derived for a dry granular fluid \cite{GD99a,L05}. On the other hand, the kinetic contributions $\eta_k$, $\kappa_k$, and $\mu_k$ are in principle functions of $\gamma$. The kinetic coefficients $\eta_k$, $\kappa_k$, and $\mu_k$ are defined, respectively, as
\beq
\label{a8}
\eta_k=-\frac{1}{(d-1)(d+2)}\int\; \dd{\bf v}\; D_{ij}({\bf V}) \mathcal{C}_{ij}({\bf V}),
\eeq
\begin{equation}
\label{a9}
\kappa_k=-\frac{1}{dT}\int\, \dd{\bf v}\; {\bf S}({\bf V})\cdot {\boldsymbol {\mathcal A}}({\bf V}),
\end{equation}
\begin{equation}
\label{a10}
\mu_k=-\frac{1}{dn}\int\, \dd{\bf v}\; {\bf S}({\bf V})\cdot {\boldsymbol {\mathcal B}}({\bf V}),
\end{equation}
where
\beq
\label{a11}
D_{ij}({\bf V})=m(V_iV_j-\frac{1}{d}V^2\delta_{ij}), \quad {\bf S}({\bf V})=\left(\frac{m}{2}V^2-\frac{d+2}{2}T\right){\bf V}.
\eeq
The coefficient $\eta_k$ can be obtained by multiplying both sides of Eq.\ \eqref{a4}  by $D_{ij}(\mathbf{V})$ and integrating over velocity. In the case of $\kappa_k$ and $\mu_k$, we multiply Eqs.\ \eqref{a2} and \eqref{a3}, respectively, by $\mathbf{S}(\mathbf{V})$ and integrate over $\mathbf{V}$. After some algebra, the expressions of the scaled coefficients $\eta_k^*$, $\kappa_k^*$ and $\mu_k^*$ can be written as
\beq
\label{a12}
\eta_k^*=\frac{1-\displaystyle{\frac{2^{d-2}}{d+2}}(1+\al)(1-3\al)\phi \chi}{\nu_\eta^*-\frac{1}{2}\left(\zeta_0^*-2\gamma^*\right)},
\eeq
\beq
\label{a13}
\kappa_k^*=\frac{d-1}{d}\left(\nu_\kappa^*-2\zeta_0^*-\gamma^*\right)^{-1}\left\{1+2a_2+3\frac{2^{d-3}}{d+2}\phi \chi
(1+\alpha)^2\left[2\alpha-1+a_2(1+\alpha)\right]\right\},
\eeq
\beqa
\label{a14}
\mu_{k}^*&=&\left(\nu_\kappa^*-\frac{3}{2}\zeta_0^*\right)^{-1}\left\{\kappa_k^*\left[2\gamma^*
+\zeta_0^{*}\left(1+\phi\partial_{\phi}
\ln \chi \right)\right]+\frac{d-1}{d}a_2\right.\nonumber\\
& & \left.+ 3\frac{2^{d-2}(d-1)}{d(d+2)}\phi \chi
(1+\alpha)\left(1+\frac{1}{2}\phi\partial_\phi\ln\chi\right)
\left[\alpha(\alpha-1)+\frac{a_2}{6}(10+2d-3\alpha+3\alpha^2)\right]\right\},
\eeqa
where $\zeta_0^*\equiv \zeta^{(0)}/\nu_0$ and we have introduced the (reduced) collision frequencies
\beq
\label{a15}
\nu_\eta^*=\frac{\int \dd{\bf v} D_{ij}({\bf V}){\cal L}{\cal C}_{ij}({\bf V})}
{\nu_0\int \dd{\bf v}D_{ij}({\bf V}){\cal C}_{ij}({\bf V})}, \quad \nu_\kappa^*=\frac{\int \dd{\bf v} {\bf S}({\bf V})\cdot {\cal L}\boldsymbol{\mathcal{A}}({\bf V})}
{\nu_0\int \dd{\bf v}{\bf S}({\bf V})\cdot \boldsymbol{\mathcal{A}}({\bf V})}, \quad
\nu_\mu^*=\frac{\int \dd{\bf v} {\bf S}({\bf V})\cdot {\cal L}\boldsymbol{\mathcal{B}}({\bf V})}
{\nu_0\int \dd{\bf v}{\bf S}({\bf V})\cdot \boldsymbol{\mathcal{B}}({\bf V})}.
\eeq
So far, the expressions \eqref{a12}--\eqref{a14} are \emph{still} exact. However, to get the explicit dependence of the reduced transport coefficients on $\al$ and $\gamma^*$, one needs to determine $\zeta_0^*$, $\nu_\eta^*$, $\nu_\kappa^*$, and $\nu_\mu^*$. As usual, the simplest approximation consists of considering the leading terms in a Sonine polynomial expansion of the zeroth-order distribution $f^{(0)}$ and the unknowns $\boldsymbol{\mathcal{A}}$, $\boldsymbol{\mathcal{B}}$, and $\mathcal{C}_{ij}$. In the case of $\zeta_0^*$, one gets Eq.\ \eqref{a16} while the final forms of the reduced collision frequencies are \cite{GSM07}
\beq
\label{a17}
\nu_\eta^*=\frac{3}{4d}\chi \left(1-\alpha+\frac{2}{3}d\right)(1+\alpha)
\left(1+\frac{7}{16}a_2\right),
\eeq
\beq
\label{a18}
\nu_\kappa^*=\nu_\mu^*=\frac{1+\alpha}{d}\chi\left[\frac{d-1}{2}+\frac{3}{16}(d+8)(1-\alpha)+
\frac{296+217d-3(160+11d)\alpha}{256}a_2\right].
\eeq

Equations \eqref{a12}--\eqref{a14} are consistent with those derived in Ref.\ \cite{GTSH12} when $\gamma^*\equiv \text{const.}$ and the Langevin-like term proportional to the parameter $\xi$ is neglected.


\end{document}